\documentclass[acmlarge]{acmart}
\makeatletter
\newcommand{\confshort}{\acmConference@shortname}
\newcommand{\conffull}{\acmConference@name}
\newcommand{\confdate}{\acmConference@date}
\newcommand{\confloc}{\acmConference@venue}
\AtBeginDocument{
  \fancypagestyle{firstpagestyle}{
    \fancyhead{}%
    \fancyfoot[C]{}%
  }
  \fancyhf{}
  \fancyhead[LO]{\@headfootfont\shorttitle}%
  \fancyhead[RE]{\@headfootfont\@shortauthors}%
  \fancyhead[LE]{\@headfootfont\footnotesize \confshort, \confdate, \confloc}%
  \fancyhead[RO]{\@headfootfont\footnotesize \confshort, \confdate, \confloc}%
  \fancyfoot[C]{}%
}
\makeatother
\acmBooktitle{\conffull\@ (\confshort), \confdate, \confloc}

\usepackage[utf8]{inputenc} 
\usepackage{enumitem}
\usepackage[T1]{fontenc}    
\usepackage{hyperref}       
\hypersetup{
    colorlinks = true,
    citecolor = blue,
    linkcolor = magenta}
\usepackage{url}            
\usepackage{booktabs}       
\usepackage{amsfonts}       
\usepackage{nicefrac}       
\usepackage{microtype}      
\usepackage{xcolor}         
\usepackage{amsmath}
\usepackage{graphicx}
\usepackage{tabularx}
\usepackage{hyperref}
\usepackage[utf8]{inputenc}
\usepackage{xcolor}
\usepackage{ifthen}
\usepackage[normalem]{ulem}
\usepackage{svg}
\usepackage{subfiles}
\usepackage{xspace}
\usepackage{wrapfig}
\usepackage{dialogue}
\usepackage{comment}
\usepackage[most]{tcolorbox}
\usepackage{subcaption}
\usepackage{booktabs}
\usepackage{multirow}

\newcommand{\1}{\mathbb{I}} 

\ifthenelse{\isundefined{\definition}}{\newtheorem{definition}{Definition}}{}
\ifthenelse{\isundefined{\assumption}}{\newtheorem{assumption}{Assumption}}{}
\ifthenelse{\isundefined{\hypothesis}}{\newtheorem{hypothesis}{Hypothesis}}{}
\ifthenelse{\isundefined{\proposition}}{\newtheorem{proposition}{Proposition}}{}
\ifthenelse{\isundefined{\theorem}}{\newtheorem{theorem}{Theorem}}{}
\ifthenelse{\isundefined{\lemma}}{\newtheorem{lemma}{Lemma}}{}
\ifthenelse{\isundefined{\corollary}}{\newtheorem{corollary}{Corollary}}{}
\ifthenelse{\isundefined{\alg}}{\newtheorem{alg}{Algorithm}}{}
\ifthenelse{\isundefined{\example}}{\newtheorem{example}{Example}}{}

\newcommand\eg{e.g.\xspace}
\newcommand\ie{i.e.\xspace}

\newcommand\tldrDone[1]{}

\newcommand\company{{pymetrics}\xspace}
  
\newcommand\biasterm{adverse impact\xspace}

\newcommand\numgames{{12 or 16}\xspace}

\newcommand\numusers{3,372,132\xspace}
\newcommand\numpositions{1,746\xspace}
\newcommand\numcompanies{156\xspace}
\newcommand\nummodels{555\xspace}
\newcommand\numapplications{4,197,168\xspace}
\newcommand\averagepositionspercompany{11.2\xspace}
\newcommand\averageapplicationsperposition{2,404\xspace}
\newcommand\averageapplicationsperapplicant{1.24\xspace}

\newcommand\totalrejectionrate{{41.8\%}\xspace}
\newcommand\totalselectionrate{{58.2\%}\xspace}

\newcommand\numsimulatedapplicants{{939}\xspace}
\newcommand\maxsimulatedmodels{{495}\xspace}

\newcommand\avgrealsimulatedmodels{{1.23}\xspace}
\newcommand\avgconnectedsimulatedmodels{{145}\xspace}

\newcommand\numsectors{{11}\xspace}
 
\newcommand\totalrevenue{{\$225 billion dollars}\xspace}

\newcommand\asianselectionrate{53.30\%}

\newcommand\whiteselectionrate{58.30\%}

\newcommand\blackselectionrate{52.50\%}

\newcommand\hispanicselectionrate{56.80\%}

\newcommand\percentAsianApplicationsToBiasedModels{{14.74\%}\xspace}
\newcommand\numberAsianApplicationsToBiasedModels{{115,317}\xspace}
\newcommand\rateAsianApplicantsToBiasedModels{{18.53\%}\xspace}
\newcommand\totalShortfallAsian{{29,320}\xspace}
\newcommand\percentageAsianBiasedPositions{{5.32\%}\xspace}

\newcommand\percentBlackApplicationsToBiasedModels{{25.87\%}\xspace}
\newcommand\numberBlackApplicationsToBiasedModels{{39,986}\xspace}
\newcommand\rateBlackApplicantsToBiasedModels{{30.70\%}\xspace}
\newcommand\totalShortfallBlack{{11,513}\xspace}
\newcommand\percentageBlackBiasedPositions{{10.62\%}\xspace}

\def\Snospace~{\S{}}

\copyrightyear{2026}
\acmYear{2026}
\setcopyright{cc}
\setcctype{by}
\acmConference[FAccT '26]{The 2026 ACM Conference on Fairness, Accountability, and Transparency}{June 25--28, 2026}{Montreal, QC, Canada}
\acmBooktitle{The 2026 ACM Conference on Fairness, Accountability, and Transparency (FAccT '26), June 25--28, 2026, Montreal, QC, Canada}
\acmDOI{10.1145/3805689.3812400}
\acmISBN{979-8-4007-2596-8/2026/06}

\begin{document}

\title{Algorithmic Monocultures in Hiring}

\author{Rishi Bommasani}
\authornote{All three authors contributed equally to this research.}
\orcid{1234-5678-9012}
\affiliation{%
  \institution{Stanford University}
  \city{Stanford}
  \state{CA}
  \country{USA}
}

\author{Sarah H. Bana}
\authornotemark[1]
\affiliation{%
  \institution{Chapman University}
  \city{Orange}
  \state{CA}
  \country{USA}}

\author{Kathleen A. Creel}
\authornotemark[1]
\affiliation{%
  \institution{Northeastern University}
  \city{Boston}
  \country{USA}
}

\author{Dan Jurafsky}
\affiliation{%
  \institution{Stanford University}
  \city{Stanford}
  \state{CA}
  \country{USA}}

\author{Percy Liang}
\affiliation{%
  \institution{Stanford University}
  \city{Stanford}
  \state{CA}
  \country{USA}}

\renewcommand{\shortauthors}{Bommasani et al.}

\begin{abstract}
Many employers screen job applicants with algorithms
built by the same few algorithm vendors. 
We hypothesize that algorithmic monoculture leads to the same individuals and members of the same racial groups facing rejection.
We acquire and analyze a novel dataset of 3 million applicants submitting 4 million applications where all the applications are screened by algorithms built by the same vendor.
We find clear racial disparities in applicant outcomes.
Of all applications submitted by Asian and Black applicants, \percentAsianApplicationsToBiasedModels and \percentBlackApplicationsToBiasedModels  
are submitted to positions that adversely impact Asian and Black applicants, respectively, according to U.S. employment discrimination standards. 
Individuals also receive homogeneous outcomes:
4\% of all applicants who apply to 10 positions are recommended for rejection from all positions, a rate higher than expected by chance.
To better understand this homogeneity, we leverage the deterministic replicability of hiring algorithms to generate the outcomes applicants would have received if they applied to all positions.  
We show that applicants would need to apply widely in order to ensure their applications are considered by a human. 
\end{abstract}

\ccsdesc[500]{Computing methodologies~Machine learning}
\ccsdesc[500]{Applied computing~Consumer products}
\ccsdesc[500]{Computing methodologies~Artificial intelligence}

\keywords{algorithmic monoculture, outcome homogenization, algorithmic fairness, algorithmic hiring, algorithmic screening}

\received{13 January 2026}

\maketitle

\section{Introduction}
Over 90\% of U.S. employers rely on hiring algorithms to screen or rank job applicants~\citep{fuller2021hidden}.  
Hiring algorithms shape which applicants are considered for an interview and which applications are never seen by a human~\citep{autor2008jobtesting, Kuhn2020,  hoffman2018discretion}. 
They are a bottleneck to opportunity \citep{Jain2024} for billions of workers.
Many employers procure hiring algorithms from the same few third-party vendors.
As of May 2023, over 60\% of the Fortune 100 and eight of the ten largest US federal agencies use HireVue's algorithms \citep{nawrat2023inside}. 
By mediating screening for multiple employers, hiring algorithms establish an  \textit{algorithmic monoculture}, defined as the state in which many decision-makers rely on the same or similar algorithms \citep{KleinbergRaghavan2021, bommasani2022picking}.  
We hypothesize that algorithmic monoculture leads to homogeneous outcomes: encountering these algorithms repeatedly will result in the \textit{same groups} experiencing \biasterm and the \textit{same individuals} being rejected~\citep{ajunwa2021, creel2022, bommasani2022picking}.

\begin{figure*}[thbp]
\includegraphics[width=\textwidth]{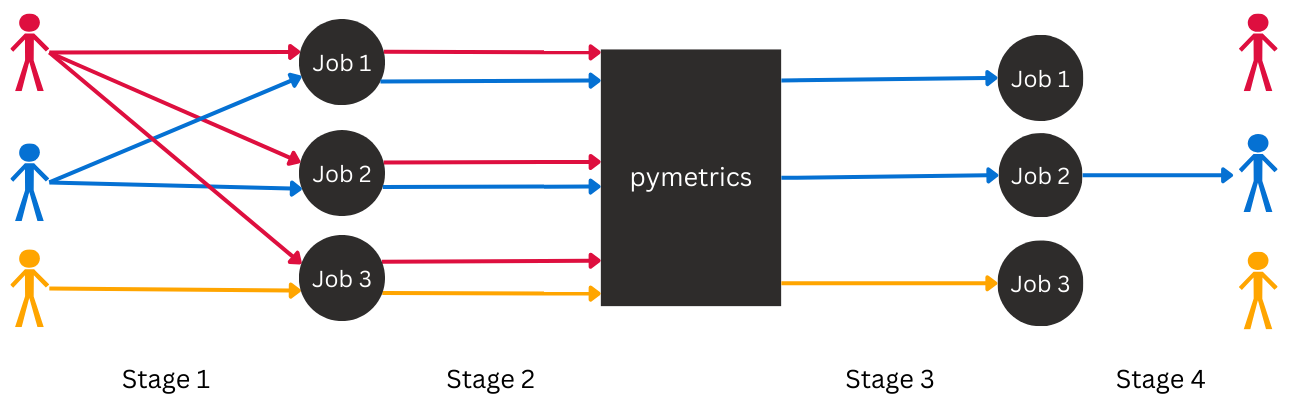}
\caption{\textbf{The \company process.}
Stage 1: Applicants apply to positions.
Stage 2: Applicants are directed to the \company platform to play assessment games.
Stage 3: \company algorithms use applicant gameplay features to recommend \totalselectionrate of applicants per position on average.
Stage 4: Employers decide which applicants to interview or hire, typically rejecting applicants that were not recommended by \company.}
\label{fig:pymetrics-diagram}
\end{figure*}

To understand algorithmic hiring in practice, we conduct the first study observing deployed algorithmic hiring decisions across multiple employers from a single vendor.
We acquire a novel dataset from the talent platform \company, which records \numapplications job applications submitted by \numusers applicants to \numpositions positions.\footnote{The data is provided for independent research: \company is unable to edit, restrict, or veto the research as stated in the data use agreement (\autoref{app:data}). This is consistent with recommendations made by \cite{Young2022CorporatePower}.}
\autoref{fig:pymetrics-diagram} shows the \company-mediated hiring pipeline.
Each application is assessed by a \company machine learning model that assigns a score that is binarized into outcomes of ``recommend'' or ``do not recommend.'' 
\company's recommendations inform their clients' decisions about which applicants to interview, reject, and hire. 
When \company's algorithms ``do not recommend'' an applicant, they are likely to be rejected without consideration by a human~\citep{fuller2021hidden}.  
Our dataset spans recommendations that influence hiring at \numcompanies employers with a cumulative annual revenue of \totalrevenue across \numsectors industries including finance, manufacturing, and warehousing.

We find evidence of  \biasterm based on race.
In a previous analysis of selection rates for many demographic groups  \citep{kassir2023hiringincontext}, researchers from \company found no differences that would prompt scrutiny according to U.S. employment discrimination law.
 Investigation into substantial differences can occur when the ``impact ratio,'' namely the ratio between the selection rates of the most selected group and the group of interest, is less than $0.8$ and statistically significant, a standard colloquially referred to as the ``4/5ths rule.''
However, this study reported only aggregate results across all applications, irrespective of the position or employer they are associated with.
Since U.S. guidelines operationalize discrimination on a per-job basis \citep[41 CFR 60-3.15.2(a)]{title41},  
we instead study each of the \numpositions positions separately.
Disaggregating on a per-position basis reveals that \percentageBlackBiasedPositions of positions demonstrate \biasterm against Black applicants. 
\rateBlackApplicantsToBiasedModels of Black applicants apply to at least one position that adversely impacts Black applicants and \percentBlackApplicationsToBiasedModels of applications submitted by Black applicants are to models with \biasterm against Black applicants. 

Our findings support the view that hiring algorithms can demonstrate \biasterm. 
Prior work finds discriminatory patterns in decisions based on applicant resumes ~\citep{Bertrand2004, Gaddis2014, Quillian2017, Thijssen2021, Quillian2023} that include demographic proxies such as racialized names like Jamal \citep{Bertrand2004} or gender-stereotyped activities like softball 
\citep{bbc2024AI}.
In contrast, \company screens applicants based on their performance on online games. Game-based and video-based assessments are among the most common and fastest-growing forms of algorithmic screening \citep{raghavan2020mitigating}.
We find \biasterm even though these games do not overtly embed demographic information and even though \company aims to pro-actively debias each model they produce~\citep[pg.~859]{kassir2023hiringincontext}.~\footnote{External researchers in a \company-funded study supported this claim, finding that the seven models they inspected were constrained to avoid \biasterm during training ~\cite[pg. 4]{wilson2021building}.}
Our findings build on prior work showing that AI can have  discriminatory effects even in the absence of explicit demographic information~\citep{BarocasSelbst2016,  caliskan2017semantics, hofmann2024ai}, as in ``proxy discrimination'' when predictive models discriminate by relying on variables that are proxies for demographic information in making their predictions~\citep{Johnson2020algorithmic, Hu2023WhatIsRace, Johnson2025proxy}.

We also provide the first evidence of systemic rejections at scale due to deployed hiring algorithms.
Prior work \citep{ajunwa2021, KleinbergRaghavan2021, bommasani2022picking, creel2022} theorizes that monoculture could lead to an applicant being rejected everywhere they apply, as is the case for the \textcolor{red}{red} applicant in \autoref{fig:pymetrics-diagram}.
42 \company models screen applicants at multiple companies, which means a rejection at one company mechanically entails rejection from another company using the same model. 
Even when applicants are evaluated by different \company models, empirically some are systemically rejected.
Of applicants that apply to ten positions, 4\% are rejected from all positions.
As applicants apply to more positions, we find that the systemic rejection rate falls exponentially ($R^2 = 0.984)$: while this exponential decay would be predicted even if decisions were made independently, the rate itself decays more slowly than would be expected by chance.
To the best of our knowledge, this pattern of systemic rejection is distinctive to algorithmic hiring. 
Analyzing data from the largest prior study of hiring decisions \cite{kline2021systemic}, which sent 83,000 applications to 108 Fortune 500 firms, we find that the systemic rejection rates observed in their data are very accurately predicted by employers making statistically independent decisions.
 Our data systematically diverges from this baseline of independence, showing that algorithmic monoculture produces qualitatively different labor market dynamics.

Data access limits empirical research on algorithmic hiring.
Single-employer studies \citep{cowgill2020bias, vandenbroeketal2021} cannot observe whether rejection at Firm A predicts rejection at Firm B because they lack cross-employer applicant tracking, correspondence studies \citep{kline2021systemic, Bertrand2004} cannot observe the underlying decisionmaking process to study the role of algorithms, and static audits of hiring algorithms \citep{wilson2021building} cannot observe deployment-time outcomes. 
Our dataset is the first to observe real algorithmic outcomes for the same applicant across multiple employers.
The algorithmic setting also allows us to study new research questions: we introduce new methods that capitalize on the deterministic replicability of algorithmic recommendations and the efficiency of algorithmic decision-making to answer a research question that would be intractable for traditional social science methods.
We generate the counterfactual outcomes applicants would receive if they applied to every position and were, thereby, assessed by every \company model. This simulation allows us to determine whether candidates are systemically rejected only because they apply to jobs for which they would be a poor fit, such that if they applied to more or different jobs they would be accepted.
Our simulation shows that every applicant would be recommended by at least one \company model, even though in reality many applicants were not recommended for any position to which they applied.
Under more realistic applicant behavior, where applicants apply more broadly but not everywhere, we find that some applicants are still systemically rejected.
To guarantee a systemic rejection rate below 0.1\%, applicants would need to submit 25 applications compared to 10 under the baseline of independence.

Independent research is necessary to illuminate otherwise-opaque hiring algorithms.
By consolidating part of hiring decision process across distinct employers, hiring algorithms impact collective adverse impact rates and patterns of systemic rejection, which are particularly salient given the established harms of discrimination and extended unemployment \citep{JacobsonLalondeandSullivan1993, sen1997inequality, sullivan2009job}.
As algorithmic hiring policy advances (\eg New York City Local Law 144 of 2021, the EU AI Act of 2024), we recommend that policymakers work to increase transparency into algorithmic hiring and create new pathways for independent research.
\section{Data}
\label{sec:data}
\noindent We analyze data from the hiring algorithm vendor \company from December 2018 through December 2022.\footnote{In August 2022, \company was acquired.}
We maintain full independence in research design, analysis, and interpretation with data access granted until December 31, 2025 subject to non-disclosure agreements regarding proprietary company information (see \autoref{app:data}). 
Our data involves job-seeking applicants, job-providing employers, and \company.
Clients of \company direct their job applicants to the \company platform, where applicants play assessment games.
\company builds hiring algorithms to make recommendations based on how applicants play the assessment games. 
Employers use \company's recommendations to make hiring decisions.

\noindent \textbf{Employers.}
Employers hire \company to use machine learning to 
classify applications.
\company's recommendations inform the employer's decisions to advance or reject applicants. 
Our data tracks \numcompanies employers that hire for \averagepositionspercompany positions on average with each position receiving \averageapplicationsperposition applications on average.
Almost 60\% of all applications are to employers in four major industries (professional services, financial services, manufacturing, and technology) for positions that are primarily white-collar with the notable exception of Hand Laborers and Material Movers (12.67\% of applications).
The majority of the \numcompanies employers are located in North America and the majority of the employers have annual revenues of at least \$5 billion.

\noindent \textbf{\company.}
\company builds 16 online games to measure applicants' cognitive traits, including propensity to take risks, processing speed, trust, altruism, and planning ability.
For each client, \company trains a binary classifier: positive training examples correspond to the gameplay features of at least 50 current employees in that role
and negative training examples correspond to the gameplay features of random profiles in the \company database~\citep{pymetricsHowWeUseAuditAI}.\footnote{Prior work has critiqued the use of game-play features for job candidate screening, including critiquing the stability of psychometric personality tests~\citep{Rhea2022}, the assumption that game-play scores can be a predictor of job success~\citep{Sloane2022HiringAlgorithms}, and the many construct reliability and validity concepts that could be used to assess the efficacy of measurement~\citep{JacobsWallach2021MeasurementFairness}. Our work does not measure the validity of the gameplay features for prediction.} Choosing which employees will serve as the positive examples is the primary way that the employer influences the classifier.  


\noindent \textbf{Applicants.}
Applicants to \company-mediated positions play either \numgames assessment games that generate high-dimensional gameplay features about the applicant. Importantly, 12 of these games are the same across all \company positions. 
These features from these games are stored and will be used again if the applicant applies to another \company-mediated position within the next 330 days.
Our data tracks \numusers applicants who each submit \averageapplicationsperapplicant applications on average to \company-mediated positions.

\noindent \textbf{Recommendations.}
When an applicant applies for a position, \company runs the associated model on the applicant's gameplay features, yielding a probabilistic score $p \in [0, 1]$.
The score $p$ is thresholded at $t = 0.5$: scores below 0.5 yield predictions $y = 0$ (``do not recommend'') and scores above 0.5 yield predictions $y = 1$ (``recommend'').\footnote{This method for converting the scores into binary outcomes is the practice \company uses when they publish results \citep{kassir2023hiringincontext} and the practice they recommended to us.
However, in practice, \company recommendations are used in different ways by different clients.
Some clients prefer the ``recommend'' category to be divided into two categories, denoted by ``recommend'' and ``highly recommend'' (sometimes color-coded as a ternary red-yellow-green system in an applicant tracking software user interface). 
Other options include quintiles where recommendations are divided into five tiers.
Throughout this work, we use the binary ``recommend'' and ``do not recommend'' categories, consistent with the analysis practices and guidance from \company.
Since our focus is primarily on rejections (\ie ``do not recommend''), the finer divisions that sometimes subdivide the ``recommend'' category from the client's perspective do not affect our analysis.}  
On average, \totalrejectionrate of applications are ``not recommended'', meaning that a hiring manager is unlikely to further consider them. 
We consider this to be equivalent to rejection in the high-volume hiring context in which \company operates. Since each position receives an average of \averageapplicationsperposition applications, it is not feasible for humans to manually review all applications.\footnote{When dealing with fewer applications, prior work finds that applicants ranked in the bottom half can be short-listed between 18--31\% of the time \citep{Fabris2026FairRanking}. For participant interviews comparing high and low-volume hiring processes that rely on algorithmic decision-making systems, see \cite{Sloane2023}.} 

\noindent \textbf{Sample.}
Our data tracks \numapplications applications. 
It includes applicant gameplay features and for each application, the application date, the position name and employer, metadata about the position and employer, and the numerical score and final recommendation each applicant received for each completed application.
40.2\% of applicants self-report race with a breakdown of 16.8\% Asian, 14.2\% White, 3.6\% Black, 3.0\% Hispanic, and all other racial categories below 2\% (\ie fewer than 100,000 applicants). 
Employers who use \company may have already collected demographic data when an applicant initially applies. 
Employers decide whether or not to allow \company to collect demographic data.
Because each applicant has a unique ID, we can track the same applicant as they apply to distinct positions across the study time period. 
The majority of applicants submit just one application to a \company-mediated position, though over five hundred thousand applicants submit multiple applications, which totals to 1.2 million applications.

\begin{table}[htbp!]
\centering
\caption{\textbf{Self-Reported Applicant Descriptive Statistics}.}
\label{tab:applicant_descriptives}
\footnotesize
\setlength{\tabcolsep}{3pt}
\begin{tabular}{@{}lr@{\hspace{0.35cm}}lr@{\hspace{0.35cm}}lr@{}}
\toprule
\multicolumn{2}{c}{\textit{Gender}} &
\multicolumn{2}{c}{\textit{Race}} &
\multicolumn{2}{c}{\textit{Country}} \\
\cmidrule(r){1-2}
\cmidrule(r){3-4}
\cmidrule(l){5-6}
Male              & 28.31 & Asian           & 16.82 & United States  & 16.05 \\
Female            & 19.47 & White           & 14.23 & India          & 6.56  \\
Prefer Not to Say & 0.17  & Black           & 3.57  & United Kingdom & 3.26  \\
Other             & 0.07  & Hispanic/Latino & 3.02  & Australia      & 2.03  \\
Missing           & 51.97 & Other           & 2.55  & Other          & 17.70 \\
                  &       & Missing         & 59.83 & Missing        & 54.39 \\
\bottomrule
\end{tabular}
\end{table}

The data is tabular with rows corresponding to applications and columns corresponding to application metadata (\eg submission time), applicant metadata (\eg race), position metadata (\eg position name), employer metadata (\eg employer name), model metadata (\eg model ID), and the percentile scores $p \in [0,1]$ that are the outputs of \company models.
As the first independent large-scale empirical study on algorithmic hiring, we elect to minimally process the data to maximize fidelity instead of smoothing away outliers or other anomalies for cleaner analyses. 
Data processing involves (i) removing test models that did not evaluate real applicants, (ii) removing unscored applications, (iii) deduplicating essentially identical applications and (iv) fixing coding errors where multiple employers were assigned the same organization ID.
These decisions had minimal effects on the data and were discussed with data scientists at \company.
\section{Results}
\label{sec:results}
We study the data using three lenses:
(i) \biasterm, a concern for hiring processes of all kinds;
(ii) systemic rejection, a particular consideration for algorithmic hiring; 
and (iii) unique possibilities enabled by algorithmic hiring.

\begin{table*}[htbp!]
\centering
\caption{\textbf{Disparate Impact Ratio Results by SOC Major Occupation Group and Race}}
\label{tab:disparate_impact}
\footnotesize
\setlength{\tabcolsep}{5pt}
\begin{tabular}{llccccc}
\toprule
SOC Major Occupation Group & Race & Impact Ratio & Positions & \#Adverse Impact & \%Adverse Impact & \%Adverse Impact-BH \\
\midrule
\multirow{4}{*}{Management (11)}
& Asian    & 0.897 & 36  & 0  & 0.0  & 0.0  \\
& Black    & 0.873 & 34  & 9  & 26.5 & 14.7 \\
& Hispanic & 0.871 & 38  & 2  & 5.3  & 0.0  \\
& White    & 0.956 & 39  & 0  & 0.0  & 0.0  \\
\midrule
\multirow{4}{*}{Business and Financial Ops (13)}
& Asian    & 0.876 & 140 & 13 & 9.3  & 7.1  \\
& Black    & 0.834 & 115 & 24 & 20.9 & 13.9 \\
& Hispanic & 0.912 & 125 & 6  & 4.8  & 0.8  \\
& White    & 0.979 & 140 & 0  & 0.0  & 0.0  \\
\midrule
\multirow{4}{*}{Computer and Mathematical (15)}
& Asian    & 0.847 & 61  & 8  & 13.1 & 11.5 \\
& Black    & 0.809 & 48  & 15 & 31.2 & 16.7 \\
& Hispanic & 0.883 & 38  & 5  & 13.2 & 2.6  \\
& White    & 0.990 & 53  & 2  & 3.8  & 1.9  \\
\midrule
\multirow{4}{*}{Architecture and Engineering (17)}
& Asian    & 0.790 & 10  & 1  & 10.0 & 10.0 \\
& Black    & 0.646 & 8   & 2  & 25.0 & 25.0 \\
& Hispanic & 0.923 & 9   & 1  & 11.1 & 0.0  \\
& White    & 0.986 & 12  & 0  & 0.0  & 0.0  \\
\midrule
\multirow{4}{*}{Legal (23)}
& Asian    & 0.816 & 12  & 1  & 8.3  & 8.3  \\
& Black    & 0.821 & 11  & 3  & 27.3 & 18.2 \\
& Hispanic & 0.912 & 12  & 0  & 0.0  & 0.0  \\
& White    & 0.859 & 12  & 0  & 0.0  & 0.0  \\
\midrule
\multirow{4}{*}{Sales and Related (41)}
& Asian    & 0.874 & 68  & 3  & 4.4  & 1.5  \\
& Black    & 0.937 & 64  & 5  & 7.8  & 3.1  \\
& Hispanic & 0.914 & 71  & 2  & 2.8  & 0.0  \\
& White    & 0.951 & 82  & 1  & 1.2  & 0.0  \\
\midrule
\multirow{4}{*}{Office and Administrative Support (43)}
& Asian    & 0.810 & 43  & 4  & 9.3  & 4.7  \\
& Black    & 0.828 & 41  & 5  & 12.2 & 7.3  \\
& Hispanic & 0.923 & 42  & 2  & 4.8  & 0.0  \\
& White    & 0.981 & 43  & 1  & 2.3  & 0.0  \\
\midrule
\multirow{4}{*}{Other SOC Code}
& Asian    & 0.927 & 28  & 3  & 10.7 & 3.6  \\
& Black    & 0.843 & 26  & 2  & 7.7  & 0.0  \\
& Hispanic & 0.921 & 23  & 0  & 0.0  & 0.0  \\
& White    & 0.906 & 27  & 0  & 0.0  & 0.0  \\
\midrule
\multirow{4}{*}{No SOC Code}
& Asian    & 0.875 & 485 & 42 & 8.7  & 4.9  \\
& Black    & 0.838 & 425 & 71 & 16.7 & 10.4 \\
& Hispanic & 0.920 & 390 & 28 & 7.2  & 1.0  \\
& White    & 0.957 & 469 & 13 & 2.8  & 0.6  \\
\midrule
\multirow{4}{*}{All}
& Asian    & 0.870 & 883 & 75  & 8.5  & \textbf{5.3}  \\
& Black    & 0.839 & 772 & 136 & 17.6 & \textbf{10.6} \\
& Hispanic & 0.916 & 748 & 46  & 6.1  & \textbf{0.8}  \\
& White    & 0.962 & 877 & 17  & 1.9  & \textbf{0.5}  \\
\bottomrule
\end{tabular}

\vspace{0.4em}
\begin{minipage}{\linewidth}
\footnotesize
For each SOC Major Occupation Group and racial group, the table reports the aggregate impact ratio across all positions, the number of positions, and the number/share of positions that demonstrate adverse impact.
We also report the share of positions that demonstrate adverse impact subject to the Benjamini--Hochberg correction (BH) threshold for $\alpha = 0.05$. 
Positions are included only if at least 30 applicants self-report race.
\end{minipage}
\end{table*}
\subsection{Adverse impact in algorithmic hiring}
We measure \biasterm according to the guidance of the U.S. Equal Employment Opportunity Commission (EEOC): the EEOC is the federal agency tasked with enforcing employment discrimination in the United States \citep{EEOC_Guidance}. 
\company acknowledges the EEOC standard as the relevant standard \citep{pymetricsHowWeUseAuditAI}.
Adverse impact occurs when there is (i) practically \textit{and} (ii) statistically significant disparities in the \textit{selection rate} $s_g$ for the group of interest $g$ when compared against the selection rate $s_{g'}$ of the most selected group $g'$.\footnote{All results for adverse impact exclusively consider applicants that self-report race.}
Practical significance requires the \textit{impact ratio} $r_g = \frac{s_g}{s_{g'}}$ to be less than $0.8$, which is why the EEOC guidance is colloquially referred to as the ``four-fifths'' rule. 
Statistical significance requires the test statistic $z_g$ to be at least $1.96$, where $z_g$ is the output of a two-sample pooled-proportion z-test with inputs $s_g$ and $s_{g'}$ along with the number of applicants for the two groups ($n_g$, $n_{g'}$).
\begin{align*}
\hat{p} &= \frac{s_g n_g + s_{g'} n_{g'}}{n_g + n_{g'}} \\
z_g &= \frac{s_g - s_{g'}}{\sqrt{ \hat{p}(1 - \hat{p}) \left( \frac{1}{n_g} + \frac{1}{n_{g'}} \right) }}
\end{align*}

For the \company data, the selection rates are: Asian (\asianselectionrate), Black (\blackselectionrate), Hispanic/Latino (\hispanicselectionrate), White (\whiteselectionrate).
The selection rates for other racial groups are lower than the White selection rate, but the disparities do not fall below the EEOC threshold since the impact ratios for all groups exceed $0.8$.
These results align with prior work from \company \citep{kassir2023hiringincontext}.
However, \company recommendations influence hiring at many distinct employers for many distinct positions. 
We argue that only analyzing impact ratios based on data aggregated from all \company-mediated positions is an improper, or at minimum an  incomplete, interpretation of the EEOC guidance, which was conceived for the purpose of identifying when a singular employer discriminates \citep[41 CFR 60-3.15.2(a)]{title41}.
Instead, when applying this standard to a hiring algorithm vendor that mediates many hiring processes, each position should be analyzed \textit{separately}.
By analogy, if only men were recommended for doctor positions and only women were recommended for nurse positions, even if the selection rates matched so as to be acceptable in aggregate, the per-position discrepancies warrant scrutiny.

When we re-conduct our analysis on a per-position basis,\footnote{For our primary results, we apply the Benjamini-Hochberg correction, which bounds the false discovery rate at $\alpha = 0.05$, because we test multiple positions for adverse impact.
However, the uncorrected results may be equally relevant in enforcing employment discrimination law since the EEOC may selectively investigate employers with positions that demonstrate adverse impact.}  disaggregation reveals \biasterm against Black and Asian applicants.\footnote{Following EEOC guidance, results are for positions with at least 30 applicants who self-report race.} 
\percentageBlackBiasedPositions of positions adversely impact Black applicants and \rateBlackApplicantsToBiasedModels of Black applicants apply to at least one of these positions.\footnote{ \citet{kline2021systemic} find that 
7\% of positions in their correspondence study discriminate against applicants with distinctively Black names.}
\percentBlackApplicationsToBiasedModels of all applications submitted by Black applicants are directed to these positions (\numberBlackApplicationsToBiasedModels applications). 
As some positions receive more applications than others, 23.3\% of positions account for 80\% of application-level adverse impact for this racial group.
We see similar effects for Asians: \percentageAsianBiasedPositions of positions demonstrate \biasterm against Asians, affecting \rateAsianApplicantsToBiasedModels of Asian applicants and \percentAsianApplicationsToBiasedModels of Asian applications (\numberAsianApplicationsToBiasedModels applications).\footnote{\company receives more Asian than Black applications, hence the discrepancy between relative and absolute.}
If applicants to the positions showing \biasterm were counterfactually recommended at the same rate as the most selected racial group, then an additional \totalShortfallBlack Black applications and \totalShortfallAsian Asian applications would have been recommended.
\autoref{tab:disparate_impact} reports the aggregate position-level \biasterm by occupational group.
Aggregating from individual positions to occupation groups suffices to mask the per-position \biasterm with the exception of Black and Asian applicants to Architecture and Engineering jobs.

We only attempt to identify \biasterm based on self-reported racial information and do not impute data, noting that we do not have access to features like applicant name.
62.35\% of all applicants in the data do not self-report race as belonging to one of the four racial groups we study.\footnote{Note also that the categories \company supplies for candidates to self-report their race reflect the U.S. Office of Management and Budget racial categories.  The options may not include categories that the respondents believe best describe them or include the terms that best reflect racial or ethnic groups in their country or region. The candidates did have the option to select ``Other''.}
We expect our results underestimate the total amount of adverse impact as a result for two reasons: more applicants from adversely impacted groups apply to positions than we count and more positions demonstrate adverse impact than we count, because positions  lacking at least 30 applicants with self-reported race are ruled out of the analysis.

\subsection{Homogeneity in algorithmic hiring}
Job seekers generally apply to multiple positions  \citep{dalton2020jobseekers}.
What happens when their applications are  screened by algorithms from the same vendor?
Prior work conjectures that an  \textit{algorithmic monoculture} \citep{KleinbergRaghavan2021} is likely to yield \textit{homogeneous outcomes}~\citep{bommasani2022picking}: some applicants are recommended for every job to which they apply and others are not recommended for any job.
The latter case, \textit{systemic rejection}, harms applicants because not being recommended by a first stage screening algorithm is very likely to result in rejection by the employer~\citep{fuller2021hidden}.

An applicant that is
``algorithmically blackballed'' \citep{ajunwa2021}
may be less likely to find a new job.
This outcome is of special concern because a  multidisciplinary literature establishes extended unemployment as harmful to individuals \citep[][\textit{inter alia}]{sen1997inequality, KroftLangeandNotowidigdo2013}.
Extended unemployment may deplete financial resources  and deteriorate both physical and mental health \citep{Rothstein2016}.
Repeated rejections may discourage future job pursuits and cause some applicants to leave the labor force altogether.
If those that are systemically rejected are unable to contribute to the labor force, 
misallocation of talent may reduce cumulative economic production \citep{hsieh2019allocation}.

\begin{figure*}[hbt!]
\centering
\begin{subfigure}{0.48\linewidth}
    \includegraphics[width=\linewidth]{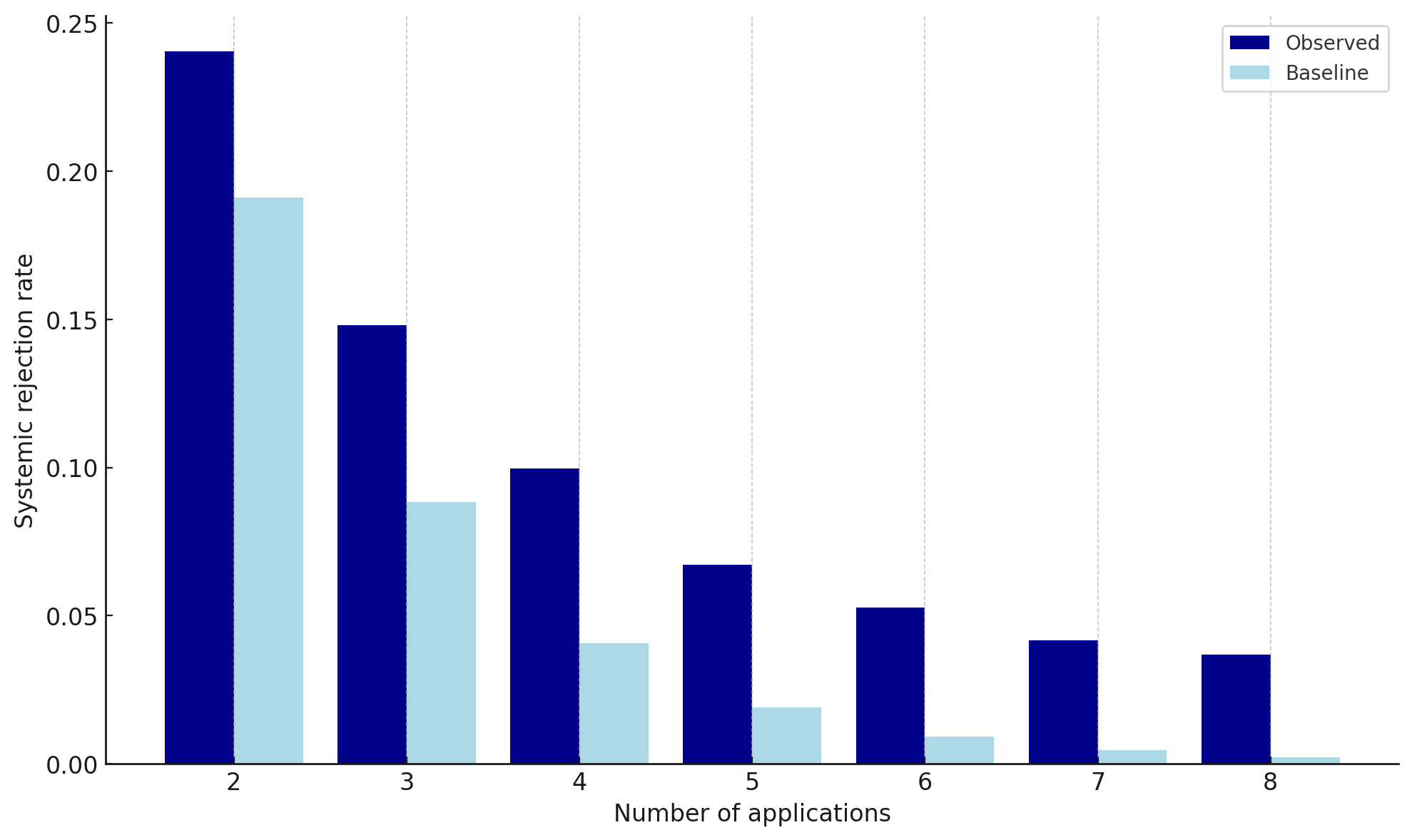}
    \caption{Data from \company}
    \label{fig:homogenization_srr}
\end{subfigure}
\hfill
\begin{subfigure}{0.48\linewidth}
    \includegraphics[width=\linewidth]{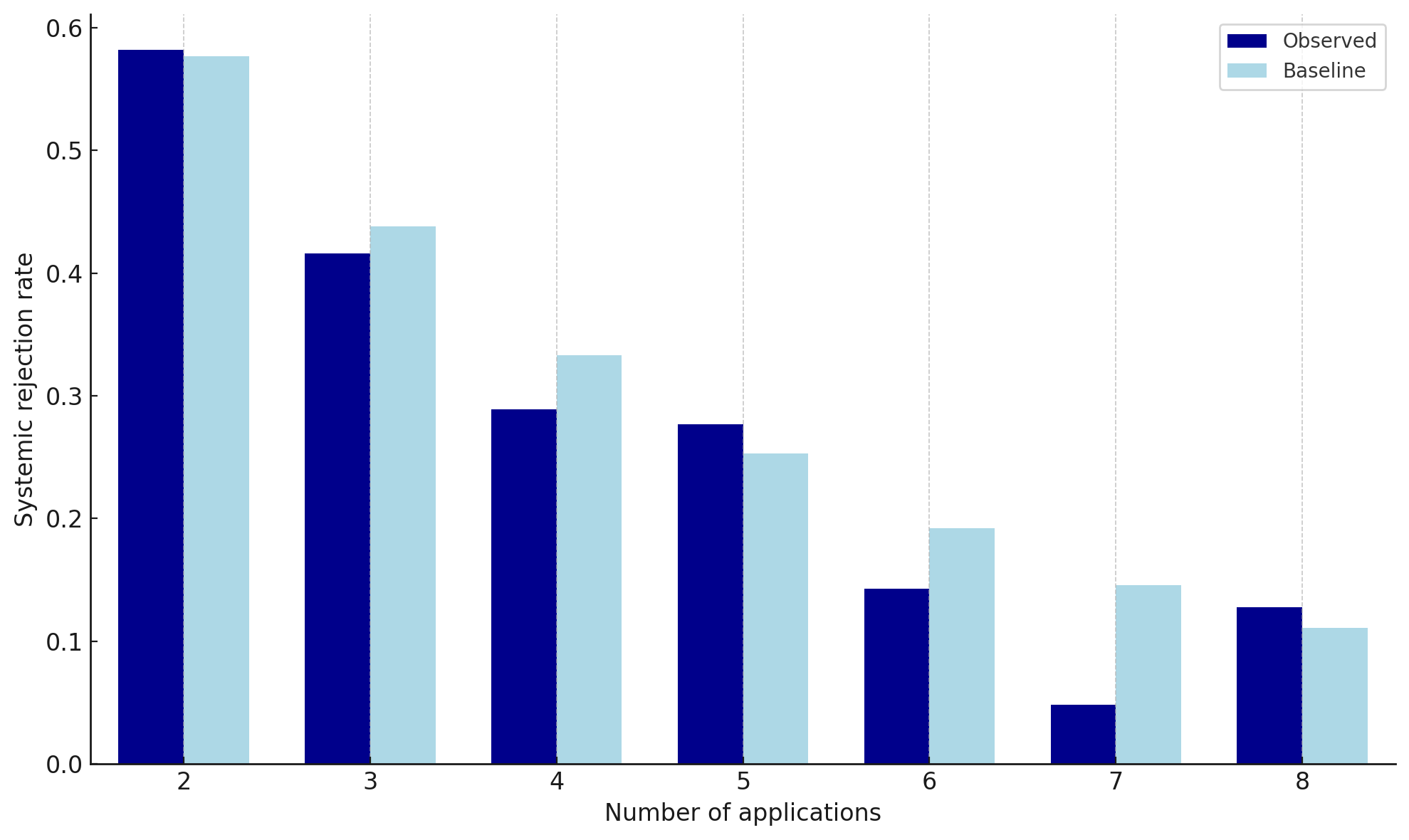}
    \caption{Data from Kline et al. (2022)  \citep{kline2021systemic}  }
    \label{fig:kline}
\end{subfigure}
\caption{
\textbf{Systemic rejection rates.}
The observed (dark blue) and baseline (light blue) systemic rejection rates for \company data (left subplot) and a correspondence study of 108 U.S. firms \citep[right subplot;][]{kline2021systemic}.}
\end{figure*}

When applying to two positions at two different employers, applicants might reasonably expect that they are receiving two separate evaluations and therefore two chances. 
But if both positions share the same model, their numerical score will be identical, which may violate  applicants' expectations of different evaluations. 
We find that 42 models are shared across positions at different companies and there are 142 unique employer pairs that share at least one model.\footnote{To identify shared models, we identify instances where the same model ID is used across different employers, though this may undercount model sharing if the same machine learning model is referred to using different model IDs.}  
Therefore, applicants in rare instances experience homogeneity mechanically due to \textit{total algorithmic monoculture} \citep{KleinbergRaghavan2021}.

While total algorithmic monoculture does occur, its impacts are limited in our dataset because very few applicants apply to positions at different employers served by the same underlying \company model. 
Therefore, we study the more general and frequent form of algorithmic monoculture \citep{bommasani2022picking}, where an applicant applies to multiple positions mediated by  \company models   (\autoref{fig:homogenization_srr}),
to understand how partial algorithmic monoculture relates to homogeneity.
For example, applicants who apply to 4 positions can receive anywhere between 0 to 4 recommendations.
10\% of these applicants are systemically rejected, as demonstrated by the observed (dark blue) bar at 4 applications.
Since applicants are only permitted to take the \company tests once every 330 days,  similar weightings of a candidate's fixed gameplay features  across different models will lead to similar outcomes for applicants across multiple applications within a year.
        
To contextualize the observed rates of systemic rejection, we distinguish systemic rejections rate foreseeable due to the underlying per-position rejection rates from further correlation between models. 
We introduce the \textit{baseline} of independence as a tool to clarify the marginal increase in systemic rejection rate due to correlated model behavior~\citep{bommasani2022picking, toups2023ecosystem}. 
In \autoref{sec:kline}, we support the validity of the baseline by showing that the largest empirical study of first-round screening procedures has a rate of systemic rejections consistent with the baseline. 
In contrast to the observed rate, which considers how often an applicant is not recommended for all $k$ positions they applied to, the baseline considers the rate at which $k$ randomly chosen applicants would not be recommended for all $k$ positions if each applicant applied to one position.  
In other words, compared to the observed systemic rate, which measures how often an applicant A is rejected from all three firms X, Y, and Z to which she applies, the baseline systemic rejection rate asks if three randomly selected applicants B, C, and D got rejected from X, Y, and Z, respectively, assuming the firms maintain their same rejection rates.


Formally, let $N$ applicants each apply to $k$ positions.\footnote{The above notation (for simplicity) assumes all applicants apply to the same $k$ positions: the general form is \autoref{app:homogeneity}.}
Define the outcome matrix $O \in \{0, 1\}^{N \times k}$ such that $O[i, j]$ is the outcome for applicant $i$ applying to position $j$, where $1$ indicates recommendation and $0$ indicates rejection.
The selection rate for position $j$ is $s_j = \frac{\sum_{i=1}^N O[i, j]}{N}$.
For $t \in \{0, \dots, k\}$, the observed rate at which applicants receive $t$ recommendations, and the baseline rate for $t$ recommendations, are defined as follows:
\begin{align}
    P_{\text{observed}}(t~ \text{rec.}) &= \frac{\sum_{i=1}^N \1 \left[t = \sum_{j=1}^k O[i,j] \right]}{N} \\
    P_{\text{baseline}}(t~ \text{rec.}) &= \text{Poisson-Binomial}(s_1, \dots, s_k)[t]
\end{align}
The observed and baseline systemic rejection rates are $P_{\text{observed}}(0)$ and $P_{\text{baseline}}(0)$, respectively.
 
For all values of $k$, we find the observed systemic rejection rate significantly exceeds the baseline rate in the \company data as shown in \autoref{fig:homogenization_srr}. 
A $\chi^2$ goodness-of-fit test strongly rejects the null that observed systemic rejection rates match the baseline ($\chi^2 = 18{,}481$, $p < 0.001$). 
Our findings establish empirical evidence to support the claim that shared dependence on a single hiring algorithm vendor yields homogeneous outcomes.

\subsection{Experimental Baseline}
\label{sec:kline}

By contrast, we find that when first round screening is not mediated by a single screening procedure, systemic rejections are close to the baseline. 
To support the empirical validity of our baseline, we study homogeneous outcomes in the largest study of first-round screening at U.S. employers to date.  Kline et al. \citep{kline2021systemic}
 generated 83000 synthetic resumes and submitted these resumes to vacant positions at 108 US companies between October 2019 and April 2021, a similar time period to our data.
The companies, which are a subset of the Fortune 500,\footnote{10 of the companies are parent companies of Fortune 500 companies.} collectively employ 15 million workers.
We analyze the homogeneity observed in the resulting callback outcomes in their data.

We find that the baseline \textit{is} an effective estimator of the  systemic rejection rate for this dataset.
 As shown in \autoref{fig:kline}, the observed systemic rejection rate is accurately predicted by the baseline  and a chi-squared goodness-of-fit test cannot reject equality of the two distributions ($\chi^2 = 20.05$, $p = 0.69$). 
In other words, while the largest previous study observes systemic rejection rates consistent with employers making statistically independent decisions, the algorithmic hiring data shows significantly correlated outcomes that lead to higher-than-baseline systemic rejection rates.
The full results are shown in the appendix in \autoref{tab:kline}.

\begin{figure*}[hbtp]
\includegraphics[width=.8\textwidth]{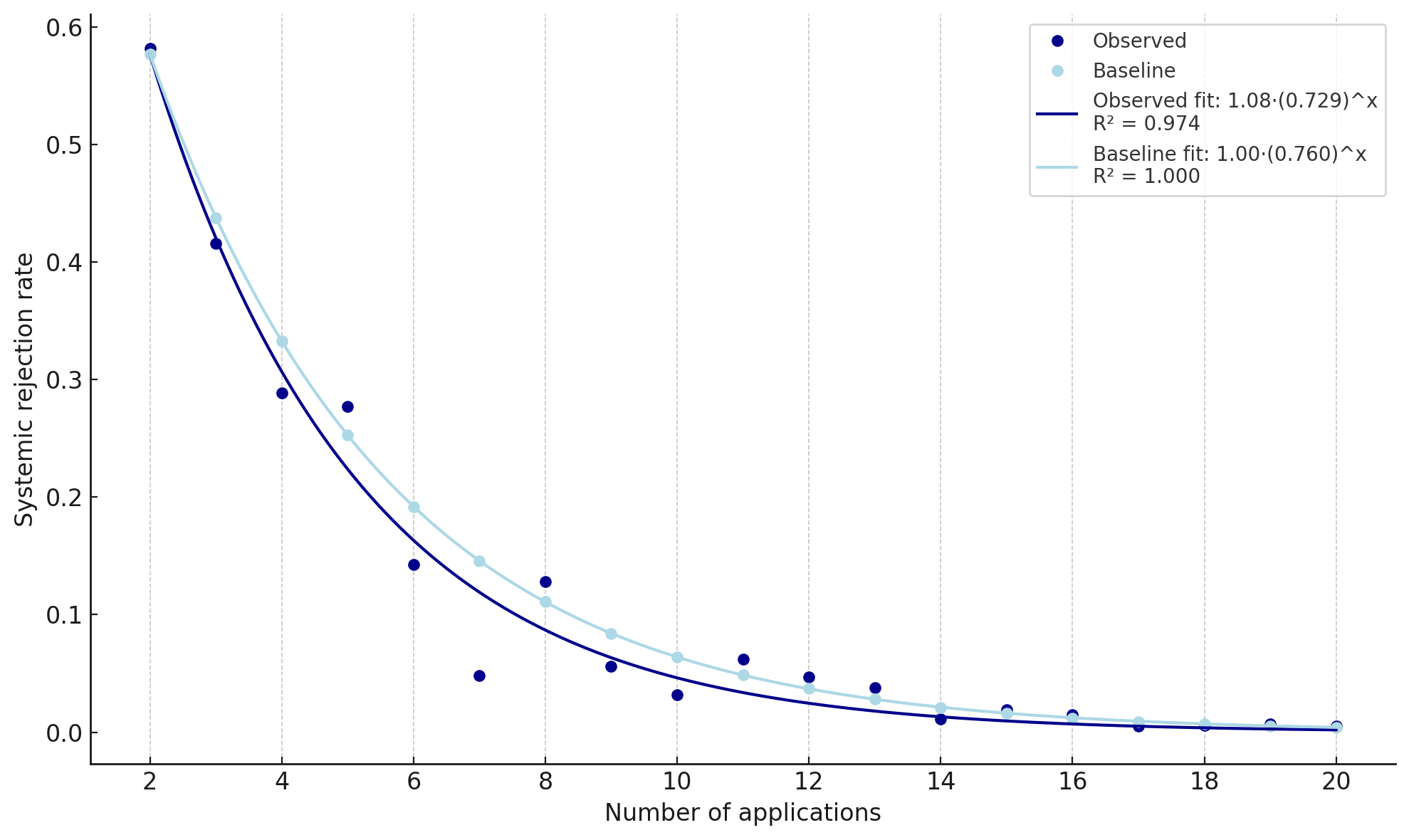}
\caption{\textbf{Kline et al. (2022) systemic rejection rates.}
We plot the observed (dark blue) and systemic rejection rates (light blue) for data from a correspondence study of 108 U.S. employers \citep{kline2021systemic} along with the  associated exponential fits.
}
\label{fig:kline}
\end{figure*}

In order to perform this analysis, we needed to decide which resumes should be considered the same. Due to \cite{kline2021systemic}'s study design, the exact same resume is generally not submitted to multiple positions.
Since we are interested in the outcomes an applicant receives when they apply to multiple positions, we treat two resumes as belonging to the same applicant if they have the same values for the following variables: firstname, lastname, race, gender, over40, associates, lgbtq\_club, political\_club, academic\_club, gender\_neutral\_pronouns, same\_gender\_pronouns. 
These variables are the targets of interest for the study. 
Other variables are chosen to make the resumes equally appealing to each employer. In particular, applicant addresses are chosen such that they are close to the location of the posted job.  
Therefore we consider a resume to be the same if it has the features listed above but a different home address.   
Under this assumption, we report the observed and baseline systemic rejection rates in \autoref{tab:kline} in the appendix. 

While both datasets involve large US employers during 2019--2020 and include entry-level positions, they differ in several ways. \citet{kline2021systemic} study callbacks---employer-initiated contact attempts---occurring at rates of 23--25\%, whereas \company assessments yield pass rates of approximately 50\%. Their sample is restricted to entry-level positions, whereas our data span entry-level through director roles. The geographic scope also differs: \citet{kline2021systemic} construct applicant profiles using addresses from public high schools within the county of each job posting, whereas our \company data reflect global hiring---6.56\% of applicants report India as their country of residence, and the most common city in our data is London. Although some \citet{kline2021systemic} employers used personality tests, \company' game-based assessments would be difficult to manipulate similarly, suggesting minimal vendor overlap. Algorithmic screening is universal in our data but likely variable in theirs: some employers used it, others did not, and those that did use it likely relied on different vendors.

Since our homogeneity measure adjusts for the underlying positive rate,  differences in callback versus pass rates do not mechanically explain the divergent findings. Nevertheless, the baseline closely tracks observed homogeneity in \citet{kline2021systemic}'s data but not in our data. This suggests that the excess homogeneity we document may be a distinctive feature of centralized algorithmic assessment systems, in which models from a single provider evaluate all applicants, rather than a general property of candidate screening processes.    

\begin{figure*}[hbt!]
    \centering
    \begin{subfigure}[b]{.27\textwidth}
        \centering  \includegraphics[width=\textwidth]{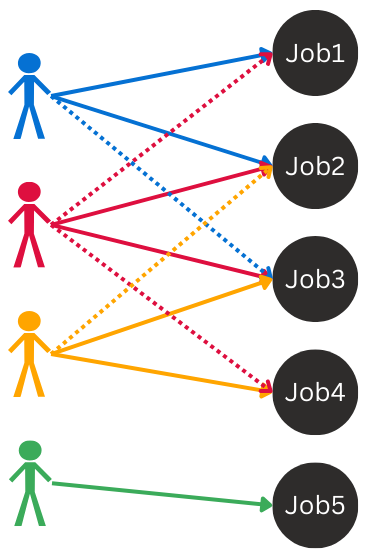}
    \end{subfigure}
    \hfill
    \begin{subfigure}[b]{.67\textwidth}
        \centering
        \includegraphics[width=\textwidth]{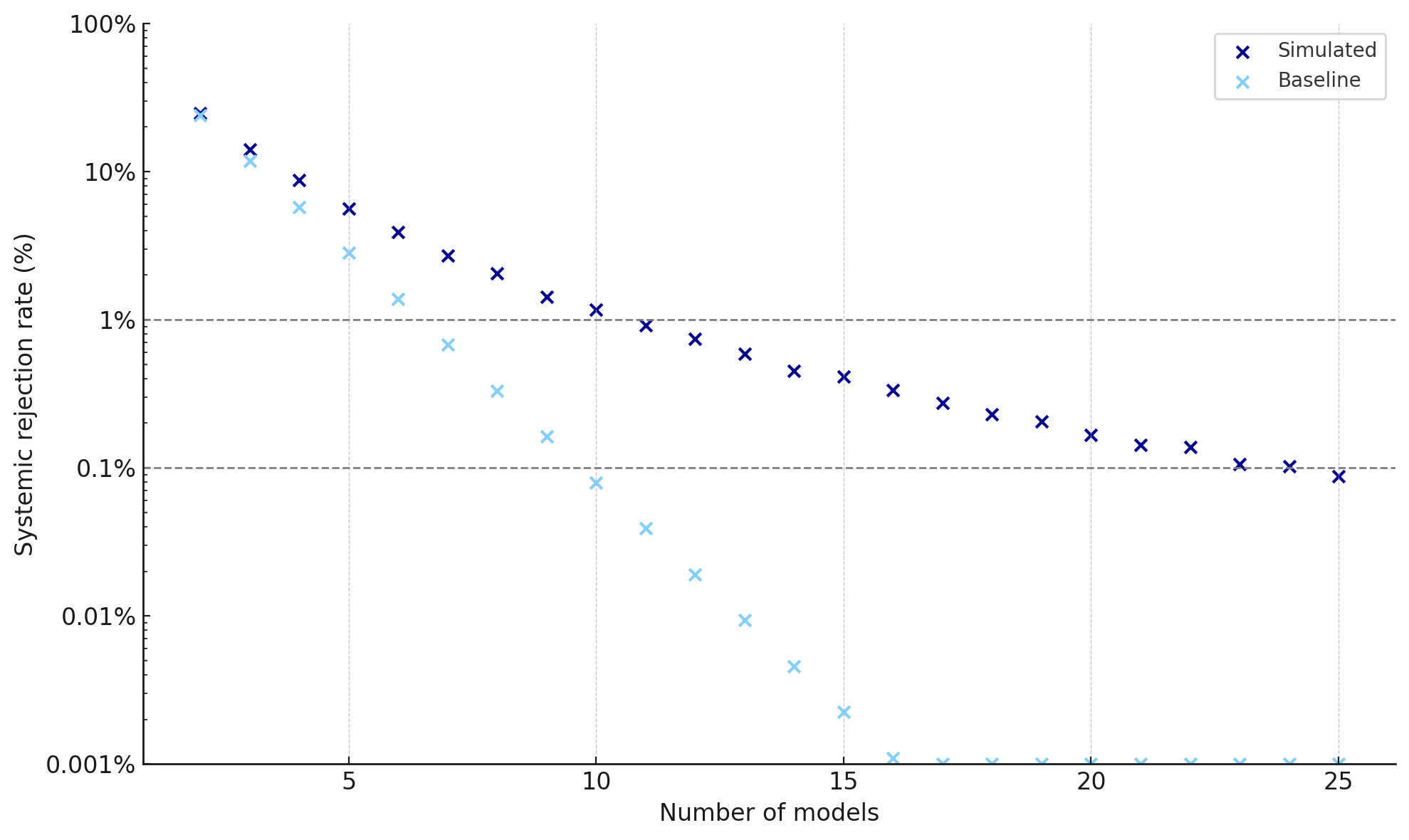}
    \end{subfigure}
\caption{
\textbf{Simulation results.}
\textbf{Left:} Given the observed set $S$ of models an applicant is assessed by (solid lines), the \textit{connected set} $S' \supseteq S$ includes every model that assessed an applicant assessed by a model in $S$ (solid and dashed lines).
\textbf{Right:} Systemic rejection rate as a function of number of models sampled from $S'$.
}
\label{fig:simulation}
\end{figure*}
\subsection{Large-scale simulation of algorithmic hiring outcomes}
\label{sec:simulation}
Our empirical findings show that applicants face both \biasterm and systemic rejection. 
Could applicants avoid systemic rejection by applying to more positions or to different positions?
Studying counterfactual outcomes is challenging for conventional methods in labor economics that rely on observational data.
In the real world, an applicant only receives an outcome if they apply to a position.
However, we leverage the deterministic replicability of \company hiring algorithms and the efficiency of algorithmic decision-making to simulate the outcomes applicants would have received if they applied to every position mediated by \company algorithms.
We sample 1000 applicants at random and ask \company to evaluate them against each of the applicable \maxsimulatedmodels models.\footnote{Since these models were developed over several years by \company, during which time they changed their infrastructure, only \numsimulatedapplicants applicants have simulated outcomes.}
Let $O_{\text{sim}} \in \{0,1\}^{939 \times \maxsimulatedmodels}$ denote the simulated binary outcome matrix where $O[i,j]$ indicates if applicant $i$ was recommended by model $j$. We find that no sampled applicant is rejected by every \company model.
The applicant who receives the fewest recommendations is still recommended by 52 models (11\%).

However, while most applicants apply to multiple positions, no applicants apply to \textit{all} positions in the labor market.
Applicants are less likely to apply to jobs in distant places, in different market sectors, or during times when they are happily employed~\citep{MarinescuandRathelot2018, Belotetal2018, KuhnandShen2023}.
Since it is unrealistic for applicants to apply to every position, we refine our analysis by studying an intermediate regime.
In this intermediate regime, applicants (i) apply more broadly than they did in reality yet (ii) do not apply to every position.
For an applicant, given the observed set $S$ of models they applied to, we define their \textit{connected set} $S' \supseteq S$ as models that share applicants with the models in $S$ (left of \autoref{fig:simulation}) 
because applicants might share preferences or constraints with applicants who applied to jobs to which the applicant also applied. 
For the sampled applicants, the average set size grows from \avgrealsimulatedmodels observed models to \avgconnectedsimulatedmodels connected models.  

Formally, let $A \in \{0,1\}^{N \times M}$ be the observed application matrix where $N$ is the total number of applicants (\ie \numusers), $M$ is the total number of models (\ie \nummodels) and $A[i,j]$ indicates if applicant $i$ was assessed by model $j$ in reality.
$B = A^TA$ is the model-model overlap matrix that encodes the number of applicants shared by every pair of models in the observed data.
Let $A_{\text{sim}} \in \{0,1\}^{939 \times \maxsimulatedmodels}$ be the submatrix of the observed application matrix $A$ corresponding to the applicants and models we simulate outcomes for.
Similarly, let $B_{\text{sim}} \in \mathbb{R}^{\maxsimulatedmodels \times \maxsimulatedmodels}$ be the submatrix of the model-model overlap matrix $B$ corresponding to the models we simulate outcomes for.
$A_{\text{sim}}' = \min(1, A_{\text{sim}}B_{\text{sim}}) \in \{0,1\}^{939 \times \maxsimulatedmodels}$ is the application matrix corresponding to our intermediate regime where applicants apply (i) to all models they applied to in reality and (ii) all models that share an applicant with those they applied to.\footnote{We describe the matrix in this way for clarity. However, condition (ii) suffices to exactly characterize the matrix, because a model must share an applicant with itself, namely the applicant being considered.}
$A_{\text{sim}}'$ expands the observed application structure yet sparsifies the fully dense structure: 
\begin{equation*}
A_{\text{sim}} \leq A_{\text{sim}}' \leq \mathbf{1}_{939 \times \maxsimulatedmodels}
\end{equation*}
\noindent 
In our analyses involving homogeneous outcomes, we prefer to study settings that fix the number of applications $k$ submitted by each applicant.
To achieve this level of control, we sample applicant outcomes so every considered applicant submits $k$ applications.
Namely, for each applicant $i$, we discard the corresponding row in $A_{\text{sim}}'$ if they applied to fewer than $k$ models (\ie $\sum A_{\text{sim}}'[i] <k$) and we sample $k$ models if they applied to at least $k$ models.

In \autoref{fig:simulation} (right), we vary the number of sampled models from each applicant's connected set of models.
Both the simulated and baseline data are well-described by exponential functions as shown by their linear scaling on the plot with a logarithmically-spaced y-axis.
The systemic rejection falls below 0.1\% first at 25 models for simulation, compared to 10 under the baseline.
That is, under the proposed applicant behavior, applicants need to apply to at least 25 different positions to ensure at least one recommendation with high probability (\ie 99.9\%). Given that the same model might be used at different positions, the number could be higher in practice. 
Since a \company recommendation only admits an applicant to the pool of applications considered by a human recruiter, applicants would need to apply to even more jobs in order to increase their likelihood of an interview.
\section{Discussion}
\label{sec:discussion}
Hiring algorithms change lives and shape labor markets, but we lack empirical research into their impact on applicants.
Most prior work centers the perspective of employers, measuring whether an algorithm reflects employer preferences, reduces employer costs, and complies with employer-level employment discrimination law.
While important, the employer-centric perspective neglects the structural shifts to the labor market caused by algorithms.
Algorithms not only influence hiring at each employer but also link outcomes \textit{across} employers due to shared dependence on the same vendor(s). 
Since an applicant's employment status depends on the \textit{cumulative decisions} of many employers, algorithmic monoculture may precipitate systemic exclusion from the labor market for some applicants.
We discuss the relationship between our work and policy (\autoref{sec:policy}) as well as key limitations relevant to the interpretation of this work and the prospect of future work (\autoref{sec:limits}).

\subsection{Policy}
\label{sec:policy}
Three types of policy bear on algorithmic hiring: (i) policies that govern hiring,  (ii) policies that govern AI/algorithmic decisions and (iii) policies that specifically govern algorithmic hiring.
In the first category, the U.S. federal law that pertains to our research on discrimination is Title VII of the Civil Rights Act of 1964. This is the origin of the impact ratio standard, or the ``4/5ths rule'', described in this paper. 
Notably, the theory of disparate-impact liability, including in the Title VII context, is subject to scrutiny under the second Trump administration's Executive Order 14281 from 2025 entitled ``Restoring Equality of Opportunity and Meritocracy''.
At the time of writing, it remains unclear how this scrutiny will affect future US employment discrimination policy.

In the second category, we highlight the European Union's AI Act and, specifically, the regulation of high-risk AI systems.
Notably, Annex III of the E.U. AI Act provides a default high-risk designation for AI systems on the E.U. market that relate to ``employment, workers’ management and access to self-employment''.\footnote{The full text identifies two categories: (a) AI systems intended to be used for the recruitment or selection of natural persons, in particular to place targeted job advertisements, to analyse and filter job applications, and to evaluate candidates and (b) AI systems intended to be used to make decisions affecting terms of work-related relationships, the promotion or termination of work-related contractual relationships, to allocate tasks based on individual behaviour or personal traits or characteristics or to monitor and evaluate the performance and behaviour of persons in such relationships.}
The systems we study likely qualify as high-risk under this designation and as of August 2, 2026, system providers and deployers will be subject to several compliance requirements.
While our measurement of adverse impact and the racial categories offered for candidate self-report by \company are both tied to U.S. law, our methods may be informative for risk management for other high-risk hiring AI systems under the E.U. AI Act.

In the third category, we highlight New York City Local Law 144, which set precedent on directly regulating algorithmic hiring.
However, since its passing in 2021, research demonstrates its limited efficacy due to issues of null compliance where employers exercise discretion over whether they are in scope of the law~\citep{wright2024null}.
In light of our findings, we stress that existing government guidance for Local Law 144 does not address the distinction we raise between aggregate and position-level impact ratios.

We hope that our research can contribute to  future evidence-based AI policy \citep{bommasani2025ebp} and therefore offer the following recommendations based on our findings.\footnote{For specificity, we refer to the U.S. context due to greater relevance of our findings and familiarity with this policy environment, but they may generalize to other jurisdictions.}

\noindent \textbf{Recommendation 1: Regulators and auditors should measure adverse impact per position.}
We find significant adverse impact at the position level that is masked in aggregate.
Since employment standards already consider position-level adverse impact, these standards should apply to measurement in the algorithmic setting, appropriately adapting to whether the data describes applications to a single position, a single employer with multiple positions, or multiple employers and positions.
Our recommendation may contradict existing guidance for New York City Local Law 144 of 2021: ``The vendor provides historical data regarding applicant selection that the vendor has collected from multiple employers to an independent auditor who will conduct a bias audit as follows:''.
The government-provided example computation of impact ratios appears to indicate all the data should be merged together, blurring distinctions between positions and even employers.\footnote{See \url{https://codelibrary.amlegal.com/codes/newyorkcity/latest/NYCrules/0-0-0-138393}.}

\noindent \textbf{Recommendation 2: Agencies should strengthen market surveillance.}
Existing authority enables agencies to partially understand employer-level hiring practices.
For example, the EEOC collects annual EEO-1 reports on employee demographics for companies with at least 100 employees.
Since employment is essential for individual welfare \citep{walzer1983, anderson2017},
there is a strong social imperative to understand homogeneous outcomes that precipitate systemic exclusion from the labor market.
If systemic rejection mediated by algorithmic hiring leads an applicant to be unemployed, or unemployed for significantly longer than they would otherwise be, the length of unemployment can be a cascading harm, decreasing the likelihood of future interviews \citep{KroftLangeNotowidigdo2013, FarberHerbstSilvermanvonWachter2019}.
Current application of agency authority is ill-equipped to address the homogeneous outcomes we find for three reasons.  First, not all systemic rejections will align with attributes that agencies are authorized to investigate because they are protected under discrimination law, such as race, age, or gender. Second, existing data collection (\eg EEO-1) is aggregated and anonymized, precluding linking of outcomes across employers.
Where legitimate privacy interests prevent even anonymized linking of outcomes, agencies should explore alternatives, perhaps through monitors for the unemployment rate and length of job searches such as the Current Population Survey (CPS).
Alternatively, akin to our work, agencies should explore applying their investigative authorities to acquire access to unique centralized datasets such as those owned by hiring vendors, which are likely to have the outcomes for the same applicants across many applications. 
Third, existing techniques to measure market concentration, while valuable, will not be an accurate measure of systemic rejection.  Models created by the same company might be more or less diverse in their rejections depending on modeling techniques, and models created by different companies can reject the same individuals, showing less diversity than might have been expected~\citep{Jain2025Rashomon}, especially when they rely on ``shared components'' such as training data, model architecture, or base model that correlate their outcomes~\citep{bommasani2022picking, kim2025correlatederrorslargelanguage}. 

\noindent \textbf{Recommendation 3: Agencies should monitor algorithmic monoculture.}
Our work demonstrates specific negative outcomes, namely racial adverse impact and systemic rejection, under the conditions of algorithmic monoculture.
Alongside measuring these risks, agencies should also consider the potential underlying structural cause.
How prevalent is algorithmic monoculture in hiring?
Awareness could anticipate harm, even if current recourse is primarily reactive, such as litigation in response to discrimination.

In addition to their potential contribution to adverse impact and systemic exclusion, the phenomena we study in this work, algorithmic hiring monocultures may be of interest for other reasons.
As with other forms of algorithmic monoculture, system-wide resilience may be compromised if hiring algorithm vendors experience outages: for example, if HireVue was unable to provide algorithmic recommendations for an extended period, hiring may be delayed or disrupted across thousands of employers including federal agencies \citep{nawrat2023inside}.
Further, other domains of employment regulation require that competitors maintain separate practices (\eg employers cannot collude to set wages).
Therefore, if competitors depend on the same hiring algorithm vendor to inform their hiring decisions, possibly by pooling data to train the vendor's algorithms, competition in hiring may be reduced to the disadvantage of applicants, as algorithmic monoculture has been shown to do in other domains~\citep{jo2025competition}.
As agencies monitor how algorithmic dependence manifests in labor markets, policymakers should consider what levels of entanglement are (un)acceptable.

\noindent \textbf{Recommendation 4: Legislators should consider whether to mandate researcher access to algorithmic hiring.}
Our work demonstrates the value of independent research, especially given the dearth of empirical research on hiring algorithms.
Empirical methods for studying hiring such as correspondence studies and surveys are effective for studying employment outcomes as well as the experiences of both applicants and employers, but struggle to isolate the effects of intermediary components such as hiring algorithms.~\footnote{Correspondence studies have also been subject to critique on the grounds of the use of deception and measurement validity \citep{Hamermesh2012, Kessleretal2019}.} 
Most prior work on algorithmic hiring relies on scraping data from publicly accessible job posting platforms like Indeed and LinkedIn \citep{zhang2022understanding}.
Research on hiring algorithms is stymied by researchers' inability to access hiring algorithms and their predictions.

\company is a noteworthy exception that has published research \citep{kassir2023hiringincontext} and provided data to external paid research collaborators \citep{wilson2021building}. 
However, 
we expect further empirical progress on hiring algorithms will be hard to come by under current conditions.
Social media researchers historically faced similar challenges: ``platforms likely will not make data available without binding legal mechanisms'' yet ``access to data for empirical research is seen as a necessary step in ensuring transparency and accountability'' \citep{nonnecke2022dsa}.
In response, the European Union's Digital Services Act of 2022 requires very large online platforms (\ie platforms with at least 45 million EU users) to provide researcher access since these platforms wield substantial and increasingly monopolistic control over information online.
Legislators should contemplate similar policy given (i) the scale of algorithmic adoption in hiring, (ii) the stakes of hiring decisions, and (iii) the absence of compelling alternatives for data-driven inquiry.

\subsection{Limitations}
\label{sec:limits}
The core limitations we identify pertain to (i) generalizability to algorithmic hiring as a whole, (ii) relationships with notions of applicant quality and model validity, (iii) relevance to the enforcement of US employment law, and (iv) certainty about downstream hiring outcomes.  In addition, we acknowledge that our work may  suppress future empirical research into algorithmic hiring and may therefore prompt stronger forms of policy response to ensure researcher access to data.

Our findings may not generalize to all algorithmic screening.
We do not have data from other vendors nor are aware of viable mechanisms to acquire such data.
In particular, the game-based approach of \company may qualitatively differ from alternative approaches such as resume screening.

We lack external measures of both applicant quality and model validity.
Because we cannot measure applicant quality, we cannot predict if applicants that were systemically rejected would have been effective at the positions to which they applied.
Lack of independent information about applicant quality affects our finding of homogenization but not our finding of \biasterm. 
No demonstration of applicant quality or fit to job is necessary to begin an investigation of employment discrimination; establishing \biasterm can be sufficient. 
We also lack external measures of model validity. We are therefore uncertain of how accurately \company models reflect the preferences of the associated employers or how successful the models are at selecting candidates who are good fits for the positions.

We cannot determine whether the evidence of adverse impact we present implies the existence of a ``less discriminatory'' algorithm or procedure that serves the same business need~\cite[78-9]{Black2024LessDiscriminatory}.
Under the relevant US law for Title VII, while evidence of adverse impact may suffice to initiate an investigation and provide probative value in litigation, it does not necessarily imply an employer's conduct is illegal. 
In particular, we do not have visibility into the development process for \company models to understand whether alternative models could have been selected or generated at similar cost.
The closest comparison we do make is to a large-scale correspondence study data~\citep{kline2021systemic}, which does not represent a single alternative screening procedure, as the acceptances and rejections were performed by a mix of human and algorithmic decision-making at many different firms. 

We do not know how the \company recommendations are used by hiring managers at employers to make final hiring decisions.  
Some prior work suggests that following the recommendations of hiring algorithms 
leads to identifying non-traditional applicants and/or better applicants~\citep{cowgill2020bias, hoffman2018discretion}; others finds that human judgments mediated by algorithmic recommendations are more discriminatory than human judgments alone~\citep{Bursell2024}. Final human decisions may be influenced by a suite of algorithmic recommendations, and different companies likely use different additional algorithmic tools in their screening processes~\citep{Sloane2022HiringAlgorithms}. 
However, we believe that not being recommended by a \company model means the downstream employer is very likely to reject the applicant, which partially abates this uncertainty in the context of (systemic) rejections.

Finally, we hope that this work encourages further independent research into algorithmic hiring, including transparency into and scrutiny of algorithmic hiring vendors.
However, our work in itself may discourage the voluntary data sharing and demographic data collection that made these findings possible.  Therefore, our work may need to serve as a foundation for policy that enables deeper inquiry into major hiring algorithm vendors.
This bears a close resemblance to the context around social media and online platform research that prompted mandatory researcher data access under the European Union's Digital Services Act~\citep{nonnecke2022dsa}.
\section{Generative AI Usage Statement}
The authors have used Generative AI tools to generate code used to generate tables, plots and macros to input the numerical values (Claude Opus 4); some authors have used it to edit portions of the text (Claude Sonnet 4). 
\begin{acks}
We thank Alondra Nelson, Anna Stansbury, Arvind Narayanan, Ashia Wilson, Bo Cowgill, Christo Wilson, Dan Ho, Deb Raji, Emma Pierson,  Erik Brynjolfsson, Ifeoma Ajunwa, Jon Kleinberg, Judy Shen, Kathleen P. Nichols, Kristian Lum, Lindsey Raymond, Manish Raghavan, Meena Jagadeesan, Mina Lee, Omer Reingold, Reva Schwartz, Rob Reich, Roger Creel, Sanmi Koyejo, Sayash Kapoor, Shibani Santurakar, Shomik Jain, Solon Barocas, Sonny Tambe, Suresh Venkatasubramanian, and Zachary Bleemer for their thoughtful feedback and guidance. We also thank Leo He, Sharmeen Malik, Frida Polli, Shea Valentine, and Georgiy Yudintsev for their support and help with accessing and understanding the data.
\end{acks}
\bibliographystyle{ACM-Reference-Format}
\bibliography{pnas}

\appendix
\newpage
\section{Data}
\label{app:data}
We acquire large-scale data from \company.
To better contextualize this data, we describe the terms that govern its use and provide greater detail into the data and its processing.

\subsection{Data terms}
We signed a Data Use Agreement (DUA) with \company that grants access to the data through December 31, 2025. The DUA imposes no restrictions on publication beyond a 30-day review period, during which \company may request removal of Company Confidential Information only. We received the data from \company through a series of secure data transfers subject to our data use agreement with \company and our data risk assessment with our research institution. The DUA also restricts redistribution of data by the researchers. 
The fully executed DUA is available at the conclusion of this appendix (with non-substantive redactions to preserve anonymity during the review process). 
The research was subject to IRB review and data risk assessment by the research institution.

\newpage 
\begin{center}
\textbf{\Large Data Access Agreement}
\end{center}

\bigskip

This Data Access Agreement (``Agreement'') is between [University] (``University''), an institution of higher education, and pymetrics Inc.\ (``Company''), a corporation, is effective on the [date] (``Effective Date'').

Company plans to provide data described as ``A database, including the following fields: anonymized candidate IDs, scores indicating candidate fit to range of pymetrics models, corresponding ONET codes for pymetrics models, demographic data collected from candidates via pymetrics' exit screen'' (``Data'') to [Principal Investigator] (``Principal Investigator''), who is a University employee, for a research project set forth in Exhibit A (``Research Program''). The parties hereby agree as follows:

\section*{GRANT AND TRANSFER}

\begin{enumerate}[label=1.\arabic*,leftmargin=2em]

\item \textbf{Grant.} Subject to the terms and conditions of this Agreement, Company grants University the nonexclusive right to use the Data solely in the Research Program.

\item \textbf{Transfer Term.} Company will make the Data available to University during the term of this Agreement, a period from: [start date] to [end date] (``Term''). The Term may be extended only by advance written agreement of both parties.

\item \textbf{No Other Rights.} This Agreement does not constitute, grant nor confer any license under any patents or other proprietary interests of one party to the other, except as explicitly stated in this Agreement.

\item Each party shall retain all right, title, and interest in its respective technology and intellectual property first conceived or reduced to practice or fixed in a tangible medium by such party before the Effective Date and independent of such party's performance of this Agreement, together with all intellectual property rights in or to the foregoing (``Background Intellectual Property''). No right, title, or interest to either party's Background Intellectual Property shall transfer to the other party under this Agreement.

\item Company will own all right, title and interest to intellectual property first conceived and reduced to practice and/or fixed in a tangible medium solely by Company's personnel that relates to Company's Data, technical information, trade secrets, know-how and any source code disclosed by Company (``Proprietary Information'') or Company's Background Intellectual Property, together with any improvements, modifications or derivative works thereof made by either party either solely or jointly (and all intellectual property rights in or to the foregoing) (``Company Intellectual Property''). University and Company will jointly own intellectual property first conceived and reduced to practice and/or fixed in a tangible medium jointly by University's and Company's personnel during the Term and directly arising from the Research Program, excluding improvements, modifications or derivative works of Proprietary Information, Company Intellectual Property, or Company Background Intellectual Property (and all intellectual property rights in or to the foregoing) (``Joint Intellectual Property''). University will own all right, title and interest to intellectual property first conceived and reduced to practice and/or fixed in a tangible medium solely by University's personnel during the term of and directly arising from the Research Project, excluding Proprietary Information, Company Background Intellectual Property, or Company Intellectual Property (or improvements, modifications or derivative works thereof and all intellectual property rights in or to the foregoing) included or incorporated therein (``Project Intellectual Property''). University hereby grants to Company a non-exclusive, royalty-free license to Project Intellectual Property, and University's rights, title, and interest in Joint Intellectual Property solely for Company's internal research and development. University will execute any documents reasonably requested by Company to effect or perfect all rights, title, and interest of Company under this section and this Agreement. Notwithstanding the foregoing, for the sake of clarity, both University and Company agree that no Joint Intellectual Property is foreseen to result from this Agreement. If there is any Joint Intellectual Property that is developed under this Agreement, Company and University will negotiate in good faith to find mutual agreement about the disposition of such Joint Intellectual Property.

\end{enumerate}

\section*{COMPANY DATA}

\begin{enumerate}[label=2.\arabic*,leftmargin=2em]

\item \textbf{Ownership.} Company retains ownership of Data. Company retains all rights to distribute the Data to other commercial or non-commercial entities. Before University's use, University shall ensure that Data shall be de-identified and shall not be used for any other purpose than those contemplated herein. University will not knowingly take any action that will enable re-identification of any Data subject.

\item \textbf{Authority.} Company warrants it has the authority to provide Data to University for use in the Research Program.

\end{enumerate}

\section*{UNIVERSITY USE OF COMPANY DATA}

\begin{enumerate}[label=3.\arabic*,leftmargin=2em]

\item \textbf{Restrictions.} University will use Data only for the Research Program as specified herein. If University seeks to use Data for other purposes, University will obtain written consent from Company, either by an amendment to this Agreement or a new agreement, before such use. University shall (1) store all Data received under this Agreement using encrypted and password protected devices; (2) maintain a list of all individuals with access to Data, as well as a log of such access, (3) restrict access to team members on a need to know basis and (4) destroy Data in all formats from all locations in both physical and digital formats, as relevant, once the purpose is achieved or the project is abandoned---whichever is earlier in time. University shall, and shall ensure Principal Investigator and its research team also shall, delete the Data within twenty-four (24) hours from the end of Term.

\item \textbf{No Further Transfer.} University will not transfer Data to any third party without the prior written consent from Company. The parties agree that no third-party processing shall be done without first executing a Data Processing Agreement.

\item \textbf{Reporting.} In consideration of Company having provided Data, University will report the results of the Research Program to Company.

\item \textbf{Compliance with Law.} University's use of Data will comply all applicable federal, state and local laws and regulations.

\end{enumerate}

\section*{CONFIDENTIAL INFORMATION}

\begin{enumerate}[label=4.\arabic*,leftmargin=2em]

\item \textbf{Definition of Confidential Information.} ``Confidential Information'' means confidential, scientific, business or financial information that is provided in written form and clearly marked as Confidential provided that such information:

\begin{enumerate}[label=(\Alph*)]
\item is not publicly known or available from other sources who are not under a confidentiality obligation to the source of the information;
\item is not already known by or available to the receiving party without a confidentiality obligation; or
\item is not independently developed by the receiving party.
\end{enumerate}

\item \textbf{No Disclosure.} The receiving party will protect the disclosed Confidential Information by using the same degree of care, but no less than a reasonable degree of care, to prevent unauthorized use or disclosure of the Confidential Information as the receiving party uses to protect its own confidential information of a like nature.

\item \textbf{Confidentiality Term.} The receiving party's obligations of confidentiality will continue for five (5) years from the date of termination or expiration of this Agreement.

\item \textbf{Compelled Disclosure.} If the receiving party is required to divulge Confidential Information either by a court of law or in order to comply with any federal, state or local law or regulation, the receiving party will provide the disclosing party with reasonable notice.

\end{enumerate}

\section*{PUBLICATION}

Principal Investigator will be free to publish and otherwise publicly disclose the Results provided that the conditions in this section and in this Agreement are fulfilled. Principal Investigator shall provide to Company a confidential copy of any such proposed publication or disclosure at least thirty (30) days prior to publication. Within that thirty (30) day review period, Company may require Principal Investigator to delete any Company Confidential Information from the proposed publication, and, if the proposed publication contains patentable subject matter directly relating to the Data, then at Company's written request within said thirty (30) day period, the Principal Investigator will delay publication for up to an additional thirty (30) days to allow for filing of patent application(s). If negative findings are to be included in Principal Investigator's final report, thirty (30) days before submission of any proposed publication or presentation or at least five (5) days before submission of any proposed abstracts, Company reserves the right to take immediate action to begin remediation efforts and have any efforts to comply with recommendations be documented in such publications, presentations and/or reports of the Results. Notwithstanding anything to the contrary herein, Company agrees to allow Principal Investigator to publish and disclose sufficient information regarding the Data to enable the complete and accurate publication of Principal Investigator's Results. University and Principal Investigator will maintain all such prepublication materials in confidence in accordance with Section 4 (``Confidential Information'') of this Agreement. The Principal Investigator will furnish Company with periodic written reports on the progress of the Research Program as mutually agreed by the parties and reasonably consistent with applicable research standards. The Results shall be formatted to meet the needs of diverse audiences, including AI researchers, I/O psychologists, HR analysts, Company's current and prospective clients, and stakeholders. Company may provide suggestions on the final format for the Results. The final draft of the Results shall include the purpose, methods and findings of the Research Program.

\section*{PUBLICITY}

Neither party will use the name or trademark of the other party, or the names of the other party's employees, students or agents in any publicity, advertising or announcement related to this Agreement without the prior written consent of the other party's authorized officials.

\section*{GENERAL PROVISIONS}

\begin{enumerate}[label=7.\arabic*,leftmargin=2em]

\item \textbf{No Warranties.} Except as stated in Section 2.2, Data are provided by Company AS IS, WITHOUT ANY WARRANTIES, EXPRESS OR IMPLIED, INCLUDING WITHOUT LIMITATION ANY WARRANTY OF FITNESS FOR A PARTICULAR PURPOSE.

\item \textbf{Liability.} In no event shall Company be liable for any use by University of Data or Results or for any loss, claim, damage, or liability, of any kind or nature, that may arise from or in connection with this Agreement or University's use, handling, or storage of Data. University agrees to indemnify and hold harmless Company, its trustees, officers, employees, students, volunteers and agents from all liability, loss, or damage they may suffer as a result of claims, demands, costs or judgments against Company arising out of the use, handling or storage of Data by University. University will ensure that Principal Investigator complies with all aspects of this Agreement and shall be responsible for any breach of this Agreement by Principal Investigator.

\item \textbf{Termination.} Either party may terminate this Agreement at any time upon thirty (30) days prior written notice, in which case University will discontinue within thirty (30) days use of the Data and related information. University agrees, upon Company's direction, to return or destroy Data. Sections 2.1, 3.1, 3.2, 3.4, 4, 5, 7.1, 7.2 will survive the termination or expiration of this Agreement.

\item \textbf{Notice.} All notices under this Agreement are deemed fully given when written, addressed, and sent as follows:

All notices to Company are e-mailed or mailed to:

\begin{quote}
pymetrics, Inc.\\
Legal Department\\
\end{quote}

All notices to University are e-mailed or mailed to:

\begin{quote}
[University Contracts Office]\\

cc: Principal Investigator
\end{quote}

\item \textbf{Severability.} If any paragraph, term, condition or provision of this Agreement is found by a court of competent jurisdiction to be invalid or unenforceable, or if any paragraph, term, condition or provision is found to violate or contravene the substantive laws of the State of [State], then the paragraph, term, condition or provision so found will be deemed severed from this Agreement, but all other paragraphs, terms, conditions and provisions will remain in full force and effect.

\item \textbf{Integration.} This Agreement, including attached Exhibits, supersedes all prior oral and written proposals and communications, if any, and sets forth the entire agreement of the parties with respect to the subject matter hereof, and may not be altered or amended except in writing and signed by an authorized representative of each party.

\item \textbf{Electronic Copy.} The parties to this document agree that a copy of the original signature (including an electronic copy) may be used for any and all purposes for which the original signature may have been used. The parties further waive any right to challenge the admissibility or authenticity of this document in a court of law based solely on the absence of an original signature.

\end{enumerate}

\bigskip

The duly authorized party representatives execute this Agreement.

\bigskip

\noindent
\begin{tabular}{p{0.45\textwidth}p{0.45\textwidth}}
\textbf{[University]} & \textbf{COMPANY} \\[1em]
Signature: & Signature: \\[1em]
Name: [Redacted] & Name: [Redacted] \\[0.5em]
Title: [Redacted] & Title: [Redacted] \\[0.5em]
Date: [Redacted] & Date: [Redacted] \\
\end{tabular}

\bigskip

I acknowledge that I have read this Agreement in its entirety and will use reasonable efforts to uphold my obligations and responsibilities under this Agreement.

\bigskip

\noindent
\textbf{PRINCIPAL INVESTIGATOR}

\noindent
Signature: \\
Name: [Redacted] \\
Title: [Redacted] \\
Date: [Redacted]

\newpage

\begin{center}
\textbf{Exhibit A -- Research Program}
\end{center}

\bigskip

Algorithmic hiring is a broadly deployed form of algorithmic decision-making. To implement algorithmic hiring, many firms rely on vendors of hiring algorithms, such as pymetrics. Given the stakes of hiring/employment, and the possible risks of algorithmic decision-making, naturally there are questions of fairness and equity: how do these systems perform across different protected categories and demographic subgroups?

In our study, we aim to further analyze the nature of the pymetrics' algorithms to see how they perform across different deployments. Concretely, our work will center on \textit{outcome homogenization}, a type of systemic harm.
That is, we will analyze whether the pymetrics algorithms have a tendency to rate the same candidates highly across all systems they would provide to clients and the same candidates lowly across all systems. We will also perform various group-level analyses in addition to this, and more generally understand a variety of individual-centric and group-centric phenomena.


\newpage

\subsection{Descriptive statistics}
In \autoref{sec:data}, we report descriptive statistics for applicant metadata (see \autoref{tab:applicant_descriptives}) along with information about \company clients and the clients' positions with more information provided in \autoref{tab:employer_descriptives}. 
In addition, we report the counts for how many applications are submitted by each distinct application as it is central to our analysis of homogeneous outcomes and systemic rejection (see \autoref{tab:application_distribution}). 



\begin{table}[t]
\centering
\caption{\textbf{Descriptive Statistics for Applications Based on Self-Reported Employer Metadata}.}
\label{tab:employer_descriptives}
\small
\setlength{\tabcolsep}{3pt}
\begin{tabularx}{\columnwidth}{@{}>{\raggedright\arraybackslash}X r@{}}
\toprule
\textbf{Category} & \textbf{\%} \\
\midrule

\multicolumn{2}{@{}l@{}}{\textit{Industries}} \\
Professional Services & 22.37 \\
Financial Services & 17.11 \\
Manufacturing & 10.53 \\
Technology & 9.21 \\
Other Industries & 40.79 \\

\addlinespace[0.5em]
\multicolumn{2}{@{}l@{}}{\textit{SOC Codes}} \\
Hand Laborers and Material Movers (53-7062) & 12.67 \\
Financial and Investment Analysts (13-2051) & 5.78 \\
Customer Service Representatives (43-4051) & 4.12 \\
Software Developers (15-1252) & 2.46 \\
Missing & 50.44 \\

\addlinespace[0.5em]
\multicolumn{2}{@{}l@{}}{\textit{Account Region}} \\
North America & 55.26 \\
Europe, the Middle East and Africa & 25.00 \\
Asia-Pacific & 17.11 \\
Missing & 2.63 \\

\addlinespace[0.5em]
\multicolumn{2}{@{}l@{}}{\textit{Client Revenue}} \\
Over \$5 bil. & 51.32 \\
\$1 bil.--\$5 bil. & 14.47 \\
\$250 mil.--\$500 mil. & 3.94 \\
\$500 mil.--\$1 bil. & 2.63 \\
Missing & 25.00 \\

\bottomrule
\end{tabularx}
\end{table}

\begin{table}[htbp!]
\centering
\caption{\textbf{Distribution of Number of Applications Across \company-Mediated Positions}.}
\label{tab:application_distribution}
\begin{tabular}{@{}lrr@{}}
\toprule
\textbf{Number of Applications} & \textbf{Count} & \textbf{Value (\%)} \\
\midrule
1   & 2,837,952 & 84.16\% \\
2   & 368,192   & 10.92\% \\
3   & 101,518   & 3.01\%  \\
4   & 35,712    & 1.06\%  \\
5   & 14,851    & 0.44\%  \\
6   & 6,631     & 0.20\%  \\
7   & 3,142     & 0.09\%  \\
8   & 1,680     & 0.05\%  \\
9   & 1,047     & 0.03\%  \\
10  & 522       & 0.02\%  \\
11  & 321       & 0.01\%  \\
12  & 201       & 0.01\%  \\
13  & 133       & 0.00\%  \\
14  & 88        & 0.00\%  \\
15  & 53        & 0.00\%  \\
16  & 30        & 0.00\%  \\
17  & 23        & 0.00\%  \\
18  & 11        & 0.00\%  \\
19  & 6         & 0.00\%  \\
20+ & 19        & 0.00\%  \\
\bottomrule
\end{tabular}
\end{table}

\section{Adverse Impact Analysis}
\begin{figure}[htbp]
 \begin{subfigure}{0.45\textwidth}
     \includegraphics[width=\textwidth]{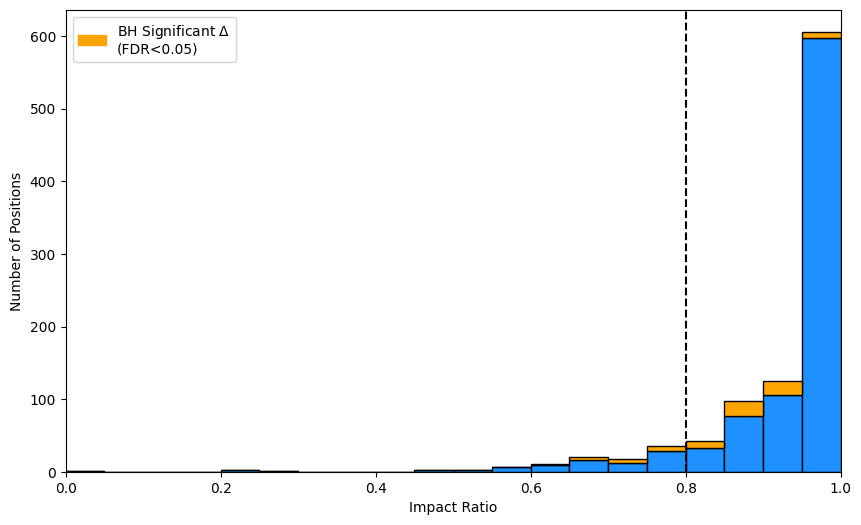}
     \caption{Female}
 \end{subfigure}
 \hfill
 \begin{subfigure}{0.45\textwidth}
     \includegraphics[width=\textwidth]{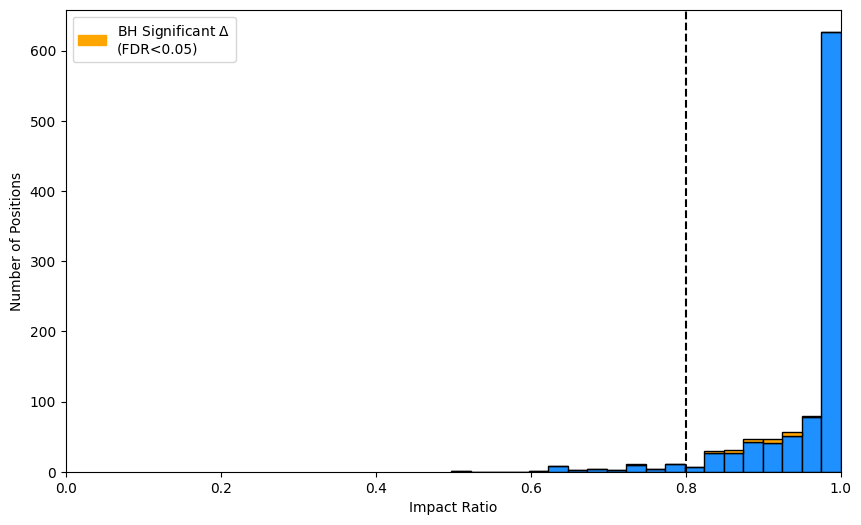}
     \caption{Male}
 \end{subfigure}
 
\caption{\textbf{Impact ratios per position by gender.}
The histograms plot the impact ratio by gender group for each of the \numpositions disaggregated positions.
We emphasize whether the impact ratio falls below $0.8$ and whether the group is selected at a rate significantly lower than that of the most select group based on a two-sample pooled-proportion z-test
for $p < 0.05$ subject to Benjamini-Hochberg correction.
If a position does not receive any applicants from members of a particular demographic group, we do not visualize the impact ratio for that position for that group.
}
\label{fig:gender_bias}
\end{figure}

\begin{figure}[p]  
\centering
\scalebox{0.9}{  
\begin{minipage}{1.0\textwidth}
\begin{subfigure}{0.45\textwidth}
    \includegraphics[width=\linewidth]{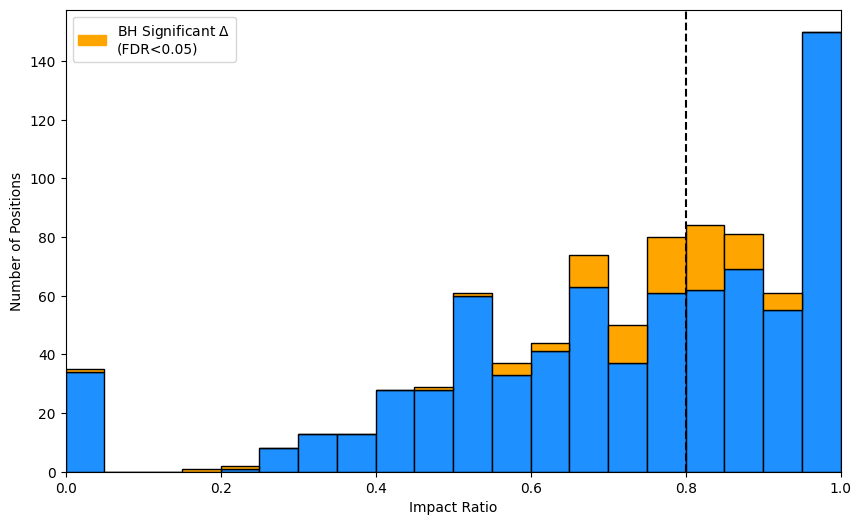}
    \caption{Female Asian}
    \label{fig:female_asian_bias}
\end{subfigure}
\hfill
\begin{subfigure}{0.45\textwidth}
    \includegraphics[width=\linewidth]{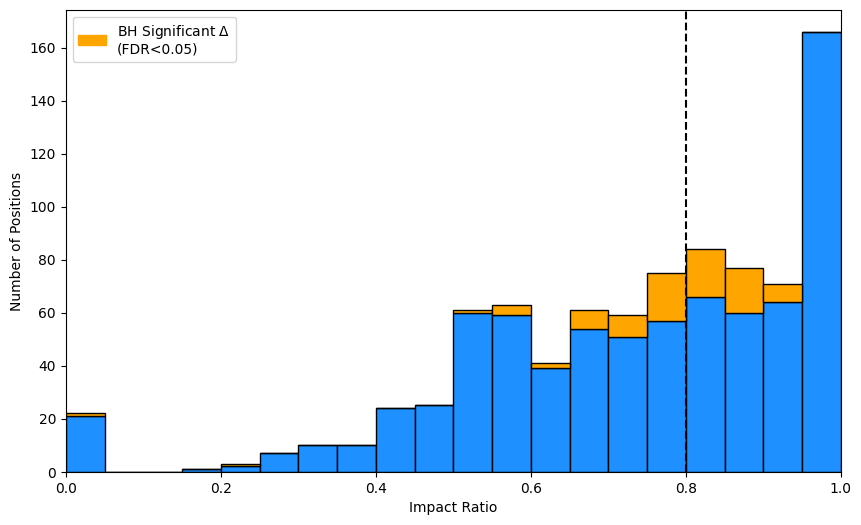}
    \caption{Male Asian}
    \label{fig:male_asian_bias}
\end{subfigure}

\vspace{0.5em}
\begin{subfigure}{0.45\textwidth}
    \includegraphics[width=\linewidth]{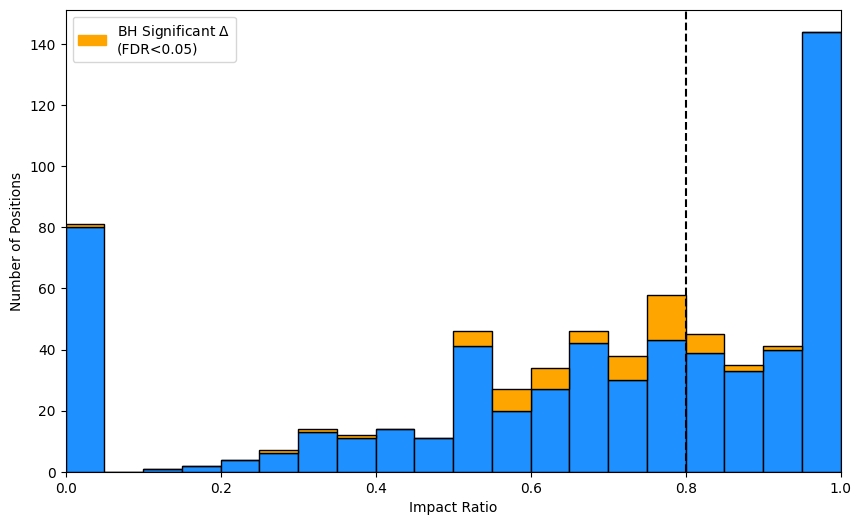}
    \caption{Female Black}
    \label{fig:female_black_bias}
\end{subfigure}
\hfill
\begin{subfigure}{0.45\textwidth}
    \includegraphics[width=\linewidth]{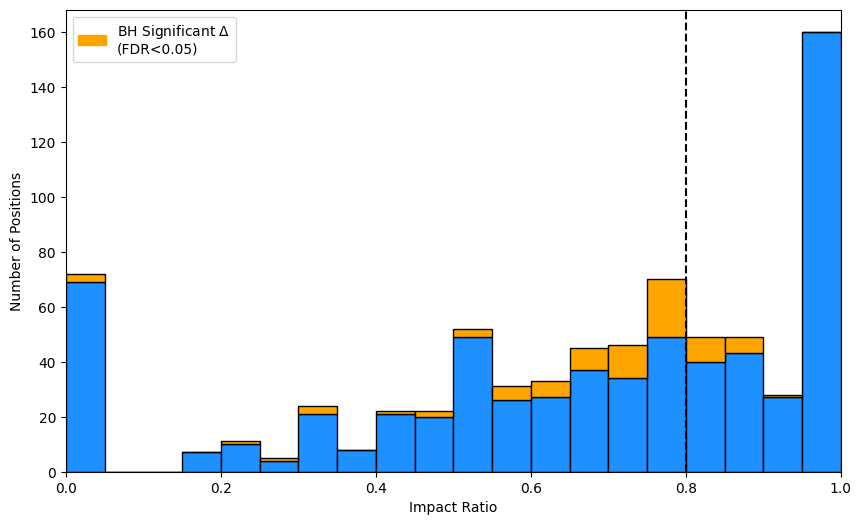}
    \caption{Male Black}
    \label{fig:male_black_bias}
\end{subfigure}

\vspace{0.5em}
\begin{subfigure}{0.45\textwidth}
    \includegraphics[width=\linewidth]{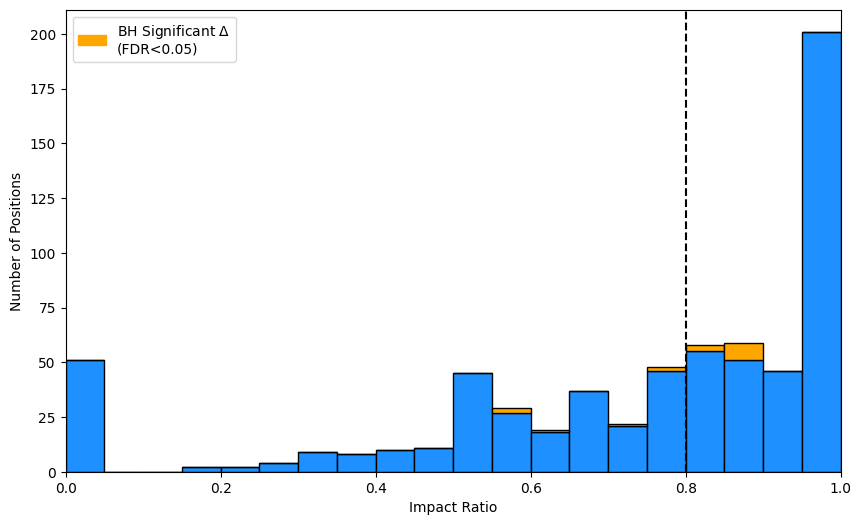}
    \caption{Female Hispanic/Latino}
    \label{fig:female_hispanic_bias}
\end{subfigure}
\hfill
\begin{subfigure}{0.45\textwidth}
    \includegraphics[width=\linewidth]{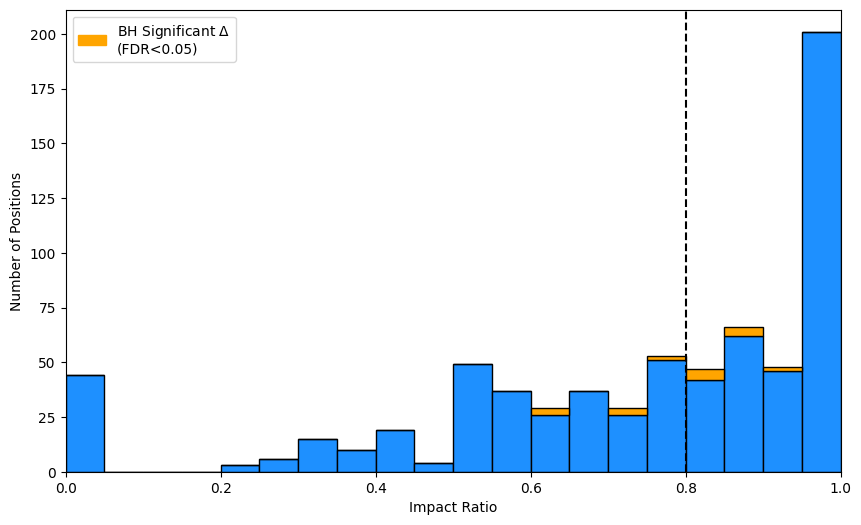}
    \caption{Male Hispanic/Latino}
    \label{fig:male_hispanic_bias}
\end{subfigure}

\vspace{0.5em}
\begin{subfigure}{0.45\textwidth}
    \includegraphics[width=\linewidth]{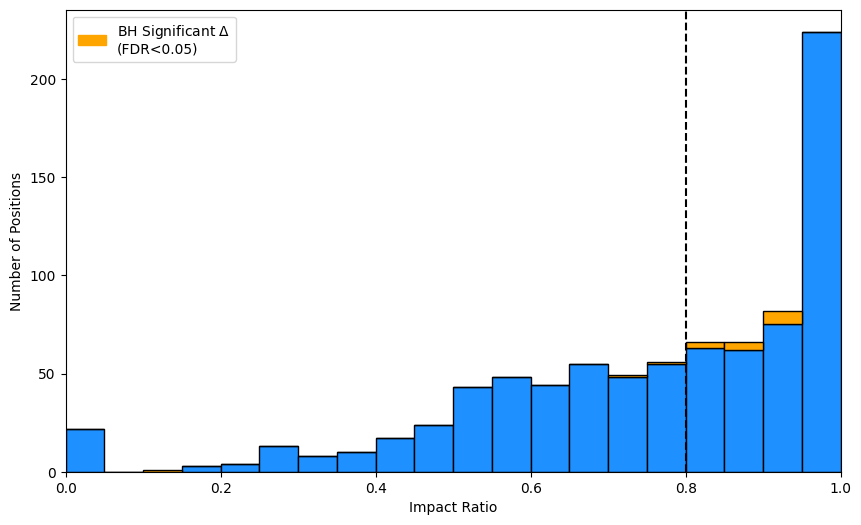}
    \caption{Female White}
    \label{fig:female_white_bias}
\end{subfigure}
\hfill
\begin{subfigure}{0.45\textwidth}
    \includegraphics[width=\linewidth]{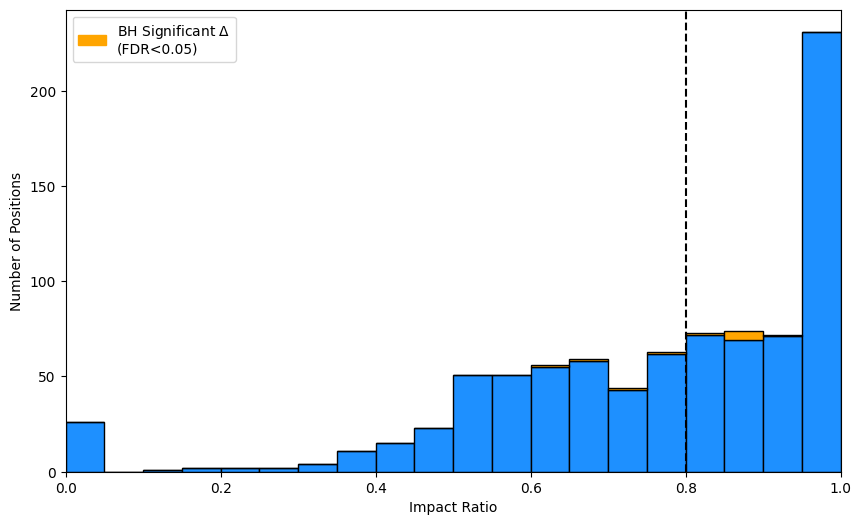}
    \caption{Male White}
    \label{fig:male_white_bias}
\end{subfigure}
\end{minipage}
} 
\caption{\textbf{Impact ratios per position by (gender, race) intersection.}
The histograms plot the impact ratio by (gender, race) intersectional group for each of the \numpositions disaggregated positions.
We emphasize whether the impact ratio falls below $0.8$ and whether the group is selected at a rate significantly lower than that of the most select group based on a two-sample pooled-proportion z-test
for $p < 0.05$ subject to Benjamini-Hochberg correction.
If a position does not receive any applicants from members of a particular demographic group, we do not visualize the impact ratio for that position for that group.}
\label{fig:intersectional_bias}
\end{figure}

We report additional empirical results on adverse impact.
In the main paper, we present the results of our adverse impact analysis when disaggregating on a per-position basis and stratifying by race and O*NET code.
In addition to those results, which we foreground because of the significant disparities for race, here we present results for gender (\autoref{fig:gender_bias}) and for gender-race intersections (\autoref{fig:intersectional_bias}).
The gender results demonstrate that very few positions meet both criteria of interest (\ie $r_g < 0.8; z_g > 1.96$) for both the Male and Female groups.
The intersectional results demonstrate similar disparities to the underlying racial groups irrespective of the gender.
More detailed statistics for all demographics groups are reported in \autoref{tab:impact_summary}.

\begin{table}[htbp!]
\centering
\caption{\textbf{Adverse impact analysis by demographic group.} 
We report the aggregate selection rate and impact ratio for each demographic group (using the microaverage across positions) as well as the number and share of models that demonstrate adverse impact against that group (Panel A).
Given the models that demonstrate adverse impact against a particular demographic group, we report the number and percentage of applications submitted to these models, the number and percentage of applicants that apply to at least one such model, and the shortfall and shortfall percentage (Panel B).
We use ``biased'' to abbreviate adverse impact.
}
\label{tab:impact_summary}
\small

\begin{tabular}{@{}lcccc@{}}
\toprule
\multicolumn{5}{c}{\textit{Panel A: Presence of Adverse Impact}} \\
\midrule
\textbf{Group} &
\begin{tabular}[c]{@{}c@{}}Aggregate\\ Selection Rate\end{tabular} &
\begin{tabular}[c]{@{}c@{}}Aggregate\\ Impact Ratio\end{tabular} &
\begin{tabular}[c]{@{}c@{}}Biased\\ Models\end{tabular} &
\begin{tabular}[c]{@{}c@{}}Biased Models\\ (\%)\end{tabular} \\
\midrule
Asian                   & 0.533 & 0.870 & 47 & 5.32 \\
Black                   & 0.525 & 0.839 & 82 & 10.62 \\
Hispanic/Latino         & 0.568 & 0.916 & 6 & 0.80 \\
White                   & 0.583 & 0.962 & 4 & 0.46 \\
Female                  & 0.551 & 0.963 & 17 & 1.75 \\
Male                    & 0.557 & 0.982 & 1 & 0.10 \\
Female Asian            & 0.519 & 0.808 & 57 & 6.70 \\
Female Black            & 0.530 & 0.803 & 56 & 8.48 \\
Female Hispanic/Latino  & 0.567 & 0.857 & 8 & 1.21 \\
Female White            & 0.582 & 0.908 & 3 & 0.36 \\
Male Asian              & 0.541 & 0.833 & 45 & 5.23 \\
Male Black              & 0.520 & 0.808 & 71 & 9.67 \\
Male Hispanic/Latino    & 0.568 & 0.874 & 9 & 1.29 \\
Male White              & 0.583 & 0.922 & 4 & 0.47 \\
\bottomrule
\end{tabular}

\vspace{1em}

\begin{tabular}{@{}lcccccc@{}}
\toprule
\multicolumn{7}{c}{\textit{Panel B: Effects of Adverse Impact}} \\
\midrule
\textbf{Group} &
\begin{tabular}[c]{@{}c@{}}Applications to\\ Biased Models\end{tabular} &
\begin{tabular}[c]{@{}c@{}}Applications to\\ Biased Models (\%)\end{tabular} &
\begin{tabular}[c]{@{}c@{}}Applicants to\\ Biased Models\end{tabular} &
\begin{tabular}[c]{@{}c@{}}Applicants to\\ Biased Models (\%)\end{tabular} &
\begin{tabular}[c]{@{}c@{}}Shortfall\end{tabular} &
\begin{tabular}[c]{@{}c@{}}Shortfall\\ (\%)\end{tabular} \\
\midrule
Asian                   & 115317 & 14.74 & 105118 & 18.53 & 29320 & 3.75  \\
Black                   & 39986 & 25.87 & 36921 & 30.70 & 11513 & 7.45  \\
Hispanic/Latino         & 2081 & 1.56 & 2063 & 2.03 & 647 & 0.48  \\
White                   & 5232 & 0.82 & 5232 & 1.09 & 1519 & 0.24  \\
Female                  & 4825 & 0.56 & 4736 & 0.72 & 1218 & 0.14  \\
Male                    & 194 & 0.02 & 194 & 0.02 & 49 & 0.00  \\
Female Asian            & 59034 & 19.61 & 54723 & 24.88 & 16602 & 5.52  \\
Female Black            & 29772 & 43.84 & 26805 & 50.78 & 7838 & 11.54  \\
Female Hispanic/Latino  & 1436 & 2.72 & 1428 & 3.46 & 427 & 0.81  \\
Female White            & 1609 & 0.67 & 1609 & 0.87 & 406 & 0.17  \\
Male Asian              & 92502 & 19.47 & 84419 & 24.61 & 24434 & 5.14  \\
Male Black              & 30158 & 35.43 & 26975 & 40.65 & 8610 & 10.12  \\
Male Hispanic/Latino    & 3421 & 4.30 & 3275 & 5.50 & 978 & 1.23  \\
Male White              & 3527 & 0.90 & 3527 & 1.21 & 1144 & 0.29  \\
\bottomrule
\end{tabular}
\end{table}

\section{Homogeneity Analysis}
\label{app:homogeneity}
We report additional information on the methods we use to study homogeneous outcomes as well as additional results.

\subsection{Methods}
In the main paper, we provide the notation for computing the observed and baseline outcomes in the setting where $N$ applicants each apply to same fixed set of $k$ positions.
Here we generalize this notation to consider $N$ applicants that each apply $k$ positions among $M$ total positions.
This framework is more general because applicants apply to the same fixed number of positions, but they may apply to entirely different positions.
Each applicant will still receive some number of recommends from $0-k$ and, as a result, the observed distribution of outcomes is computed in the same way.
However, because the positions vary across applicants, the relevant baseline distribution will vary across applicants to reflect the specific positions they applied to.
This generalization extends the prior work \citep{toups2023ecosystem} that did not need to consider this complexity.

Define the application matrix $A \in \{0, 1\}^{N \times M}$ such that $A[i, j]$ indicates if applicant $i$ applied to position $j$ where $1$ indicates they applied and $0$ indicates they did not apply.
By definition, $\sum_{j=1}^MA[i,j] = k$ for all $i$.
Define the outcome matrix $O \in \{0, 1\}^{N \times M}$ such that $O[i, j]$ is the outcome for applicant $i$ applying to position $j$, where $1$ indicates recommendation and $0$ indicates rejection.
$O$ is not defined where the given applicant did not apply to the given position.
For $t \in \{0, \dots, k\}$, the observed rate at which applicants receive $t$ recommendations, and the baseline rate for $t$ recommendations, are defined as follows:
\begin{align}
    P_{\text{observed}}(t~ \text{rec.}) &= \frac{\sum_{i=1}^N \1 \left[t = \sum_{j=1}^M 
    A[i, j]O[i,j] \right]}{N} \\
    P_{\text{baseline}}(t~ \text{rec.}) &= \frac{\sum_{i=1}^N\text{Poisson-Binomial}(\{s_j | A[i,j]=1\})[t]}{N}
\end{align}

\subsection{Figures on Homogenization} 

The following figures demonstrate the observed and baseline number of model rejections based on the \company data. 

\begin{figure}[htbp]
    \centering
    \includegraphics[width=0.32\textwidth]{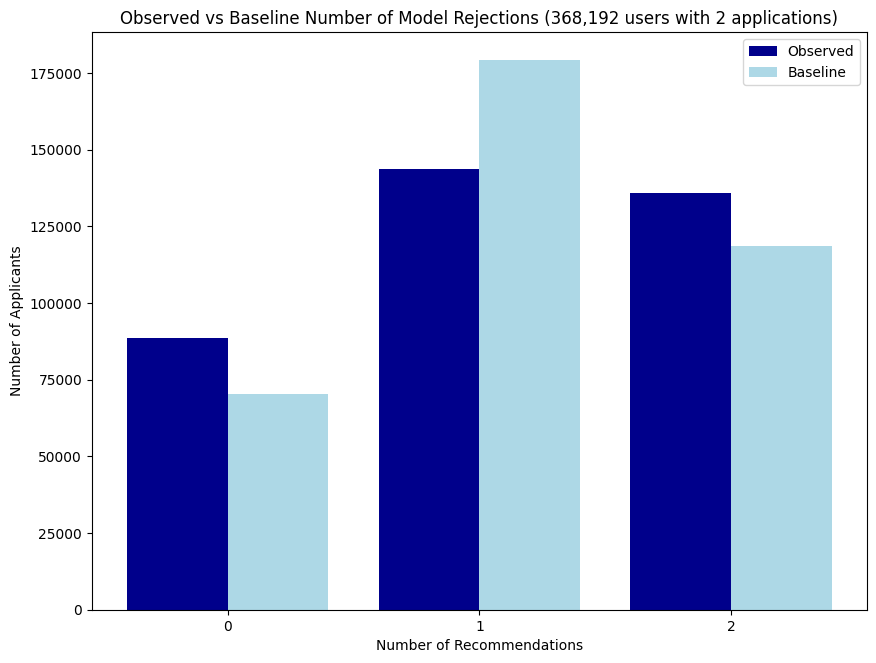}
    \includegraphics[width=0.32\textwidth]{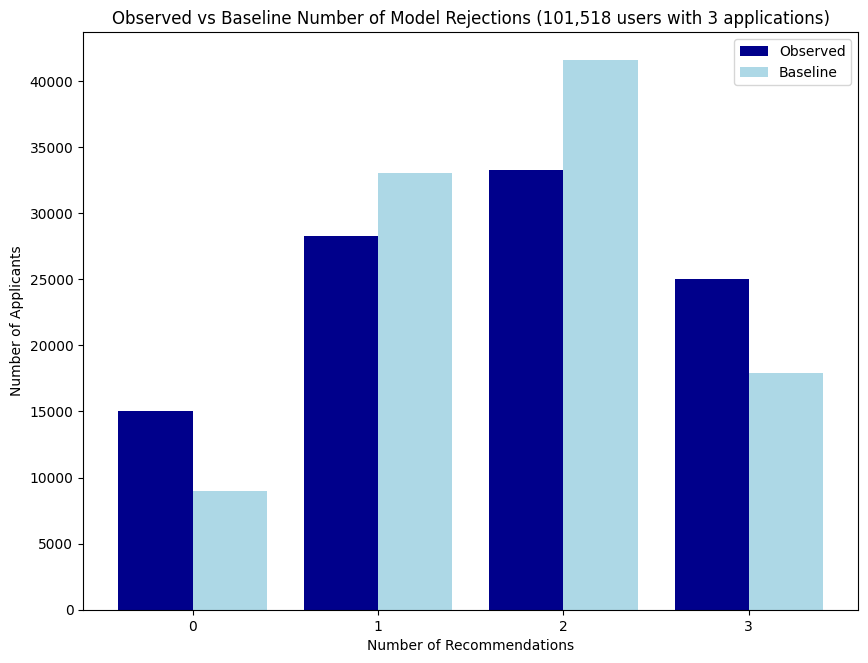}
    \includegraphics[width=0.32\textwidth]{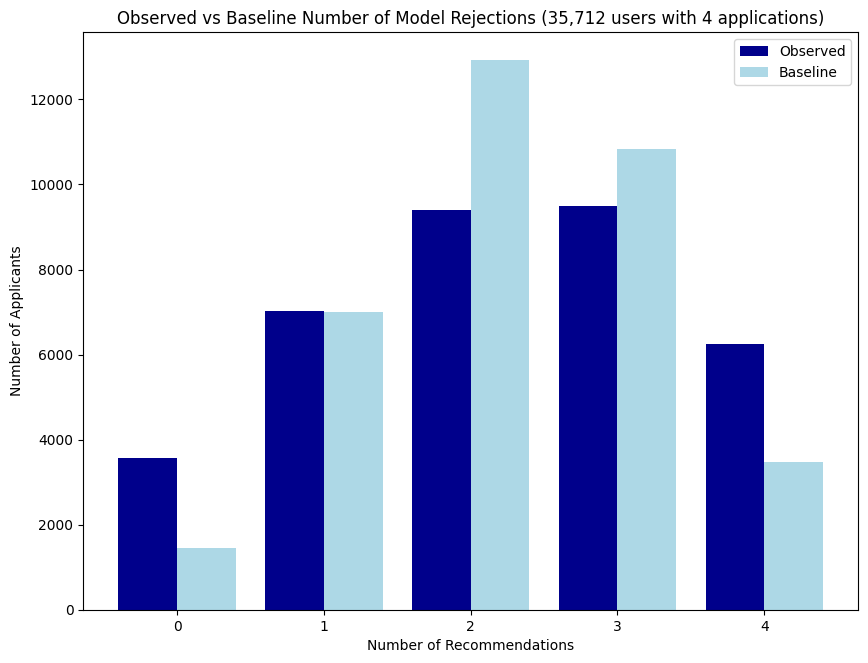}\\[0.5em]
    \includegraphics[width=0.32\textwidth]{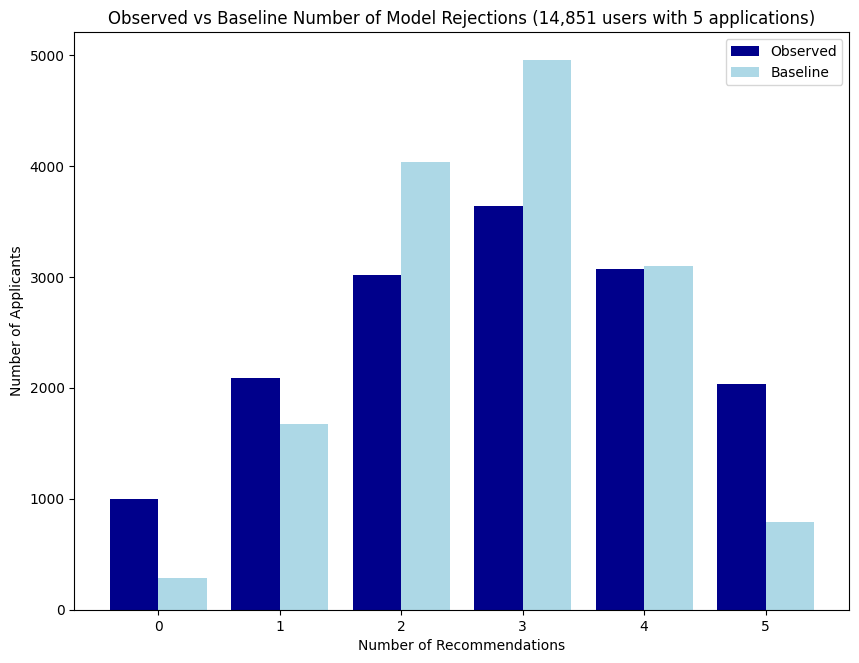}
    \includegraphics[width=0.32\textwidth]{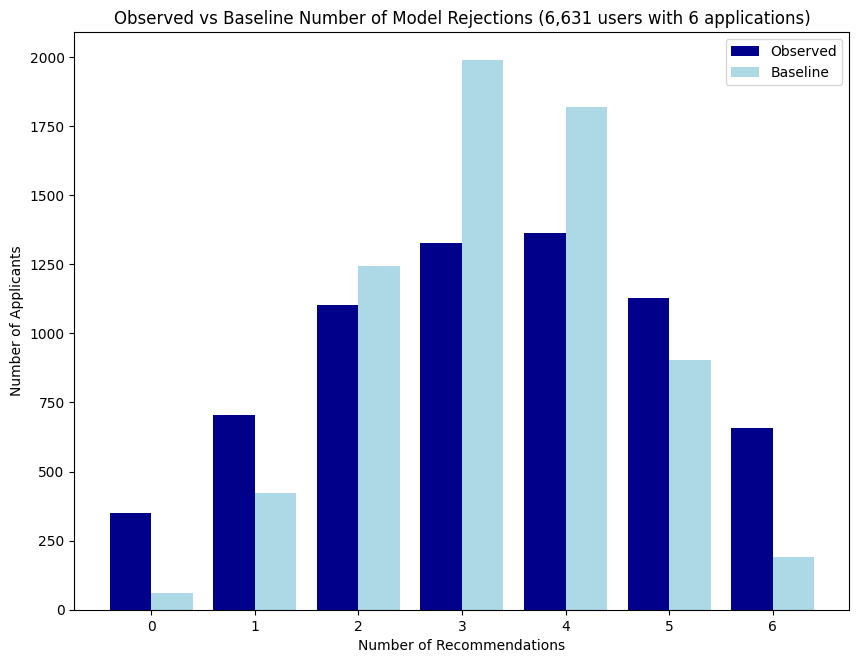}
    \includegraphics[width=0.32\textwidth]{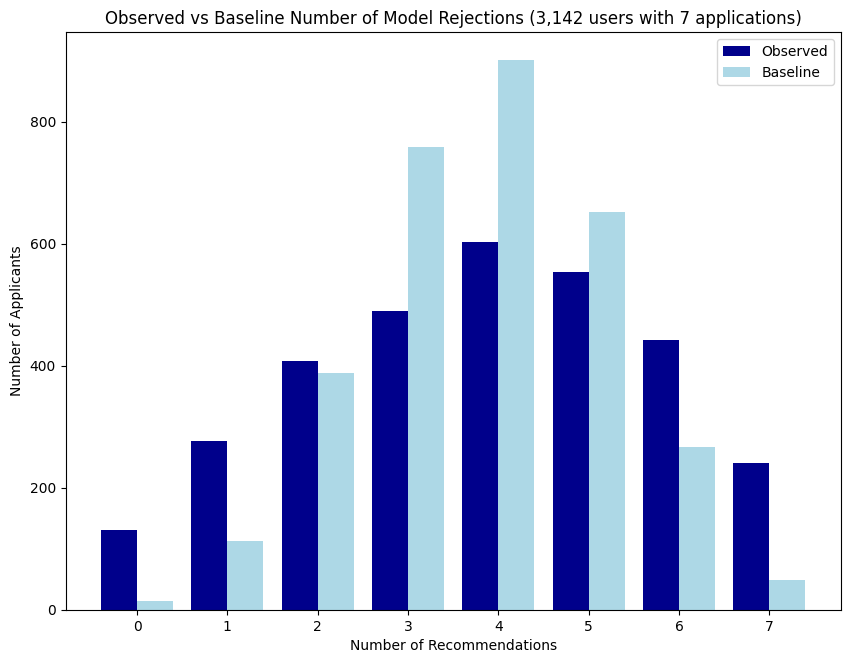}\\[0.5em]
    \includegraphics[width=0.32\textwidth]{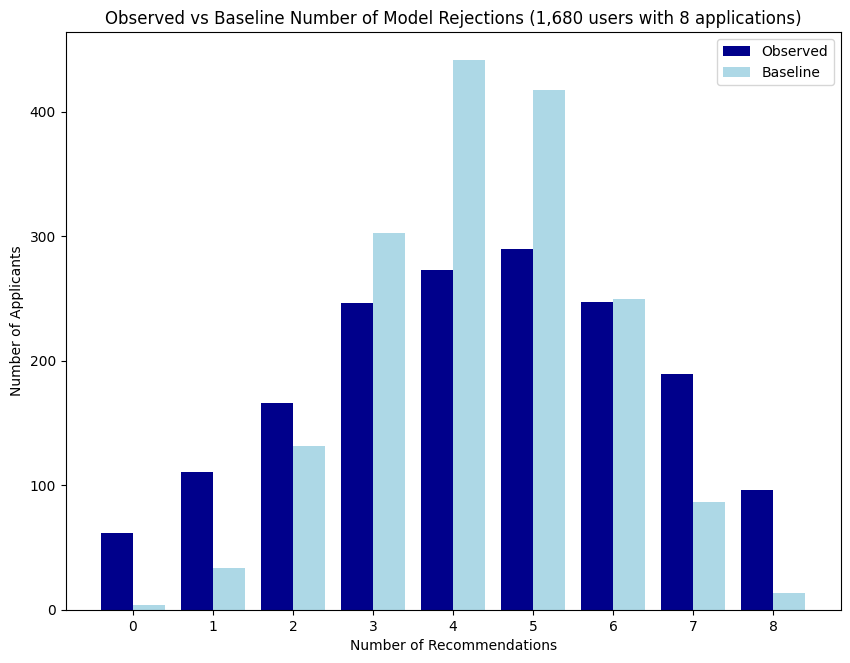}
    \includegraphics[width=0.32\textwidth]{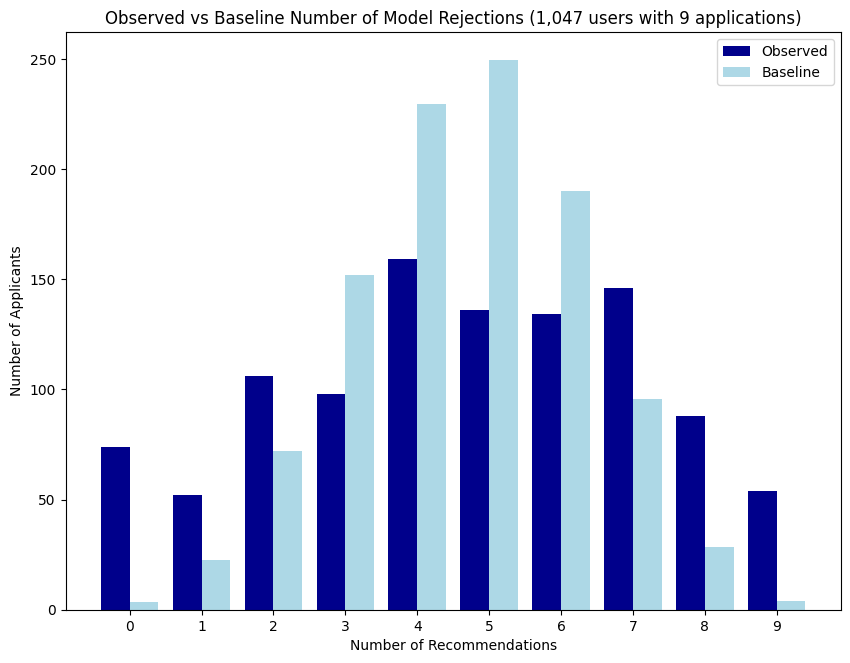}
    \includegraphics[width=0.32\textwidth]{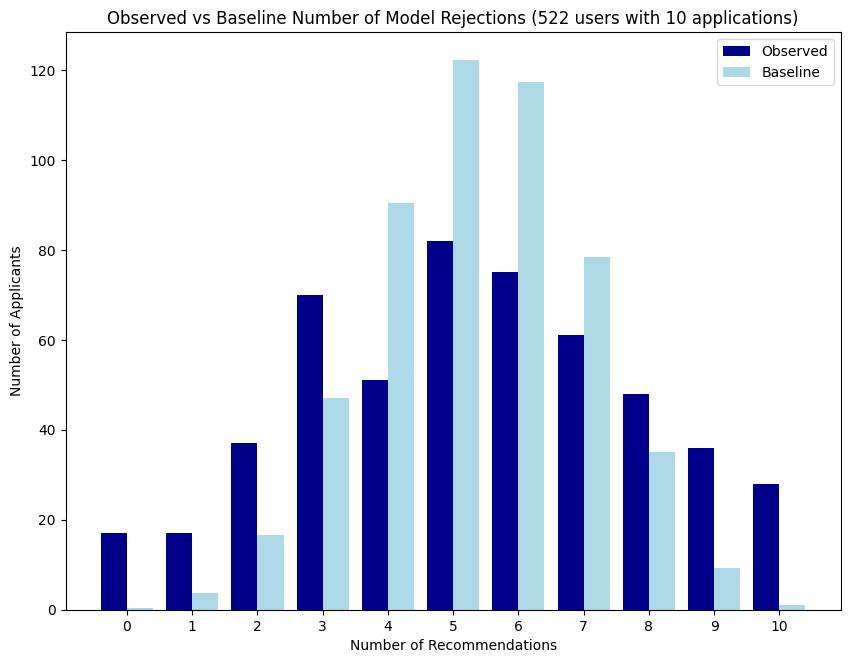}\\[0.5em]
    \includegraphics[width=0.32\textwidth]{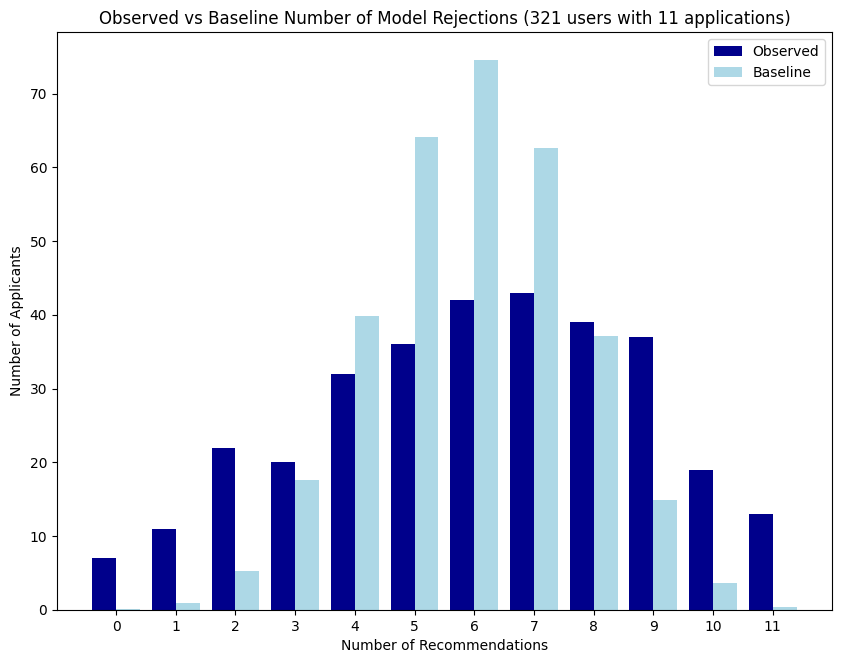}
    \includegraphics[width=0.32\textwidth]{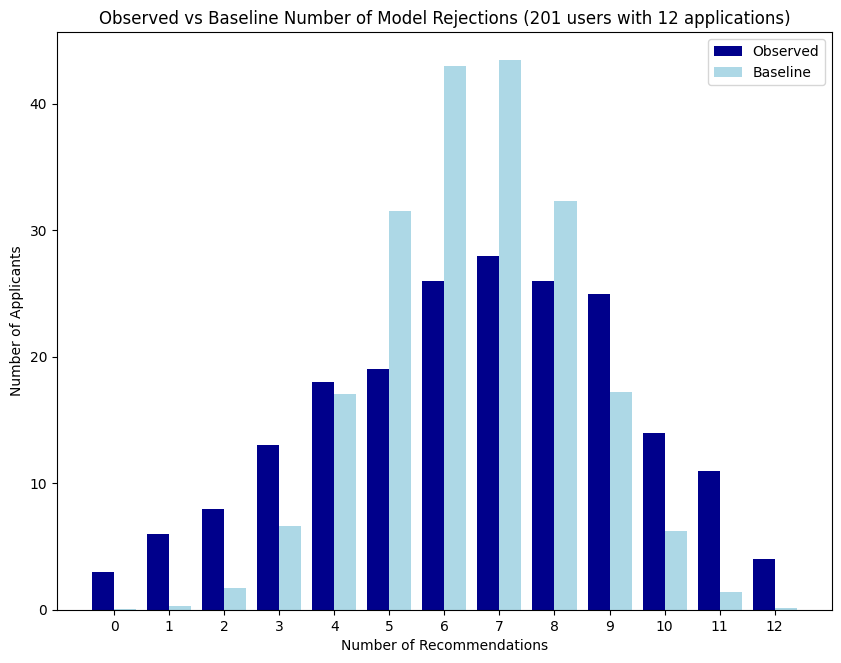}
    \includegraphics[width=0.32\textwidth]{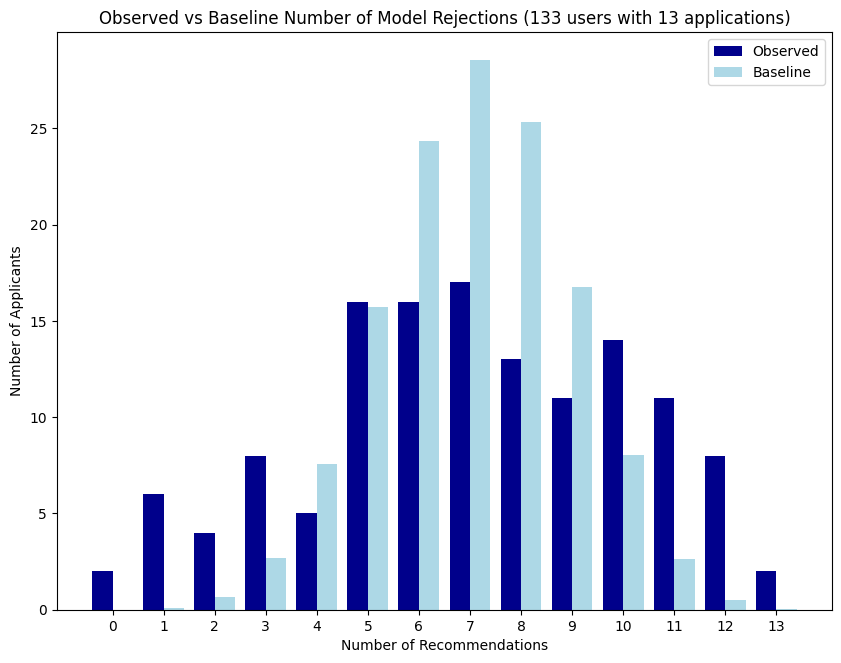}
\caption{\textbf{Homogeneous outcomes by number of applications.} 
Subfigures depict the outcomes for applicants that submit exactly 2 through 13 applications: applicants are considerably more likely to receive homogeneous outcomes than under the baseline.}    
\label{fig:homog}
\end{figure}

The following figure shows the homogeneous outcomes for the simulation exercise in \autoref{sec:simulation}. 

\begin{figure}[htbp]
    \centering
    \includegraphics[width=0.8\textwidth]{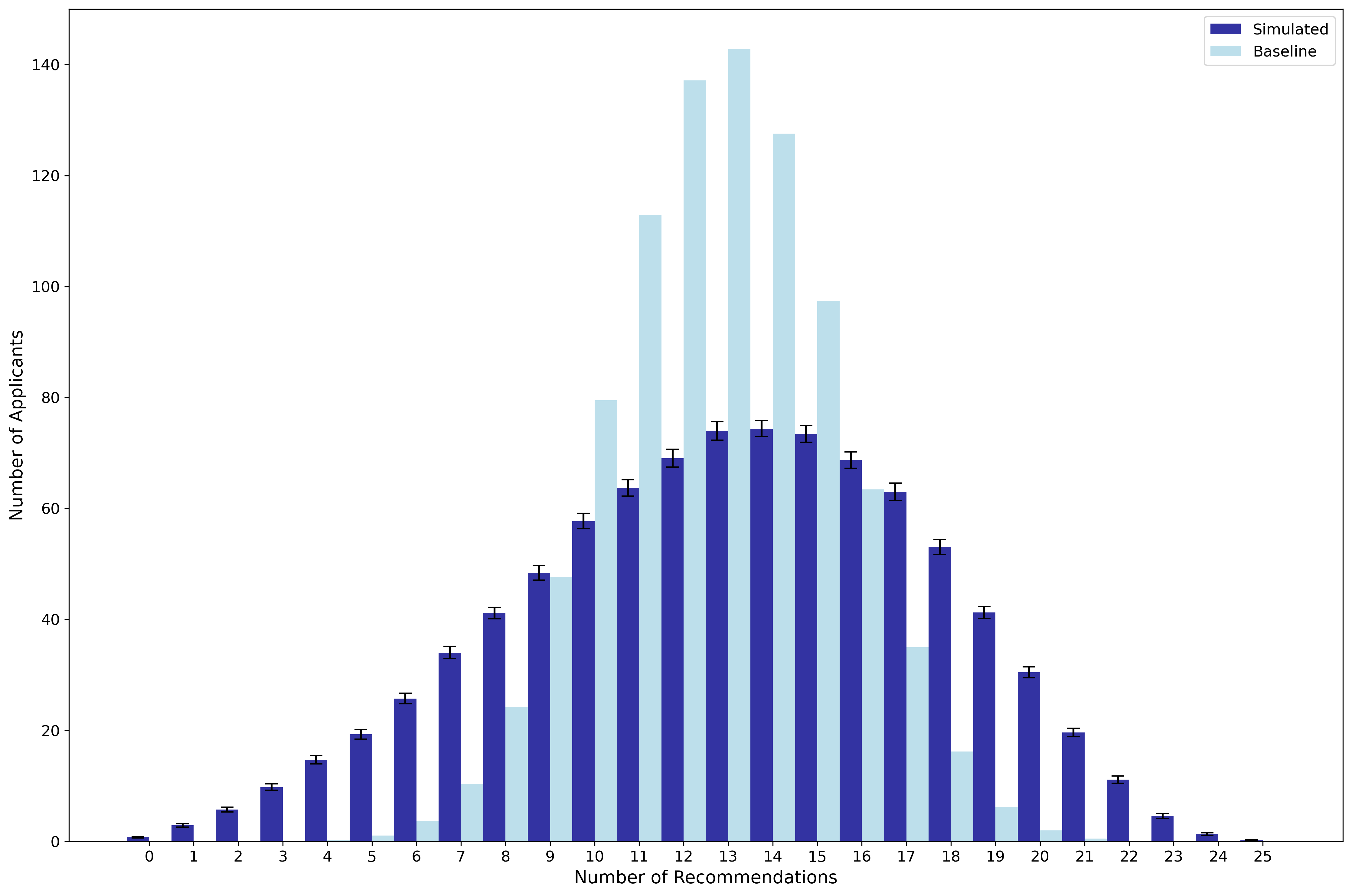}
    \caption{\textbf{Observed vs Baseline Number of Model Rejections in the Connected Set} Given the observed set $S$ of models an applicant is assessed by in reality, the \textit{connected set} $S' \supseteq S$ contains every model that assessed an applicant also assessed by a model in $S$. Applicant outcomes for 25 randomly sampled models in $S'$. Confidence intervals are based on 100 simulations.}
    \label{fig:connected_set_rejections}
    
\end{figure}

\subsection{Additional results from Kline et al.}
Beyond studying \company data, we also study homogeneous outcomes in data from Kline et al. \citep{kline2021systemic}.
In \autoref{tab:kline} we report the underlying observed and baseline systemic rejection rates that we visualize in \autoref{fig:kline}.

\begin{table}[t]
\centering
\caption{\textbf{Homogeneous outcomes in Kline et al. (2022) data.}
The observed and baseline systemic rejection rates in the Kline et al. correspondence study \citep{kline2021systemic} involving 108 US companies.
}
\label{tab:kline}
\resizebox{0.4\linewidth}{!}{
\begin{tabular}{l r r r}
\toprule
Number of applications & Count & Baseline & Observed \\
\midrule
1  & 8,209  & 76.0\% & 75.5\% \\
2  & 3,105  & 57.7\% & 58.2\% \\
3  & 1,070  & 43.8\% & 41.6\% \\
4  &   318  & 33.3\% & 28.9\% \\
5  &    65  & 25.3\% & 27.7\% \\
6  &    21  & 19.2\% & 14.3\% \\
7  &    21  & 14.6\% & 4.8\%  \\
8  &    39  & 11.1\% & 12.8\% \\
9  &    71  & 8.4\%  & 5.6\%  \\
10 &   124  & 6.4\%  & 3.2\%  \\
11 &   193  & 4.9\%  & 6.2\%  \\
12 &   279  & 3.7\%  & 4.7\%  \\
13 &   346  & 2.8\%  & 3.8\%  \\
14 &   355  & 2.1\%  & 1.1\%  \\
15 &   375  & 1.6\%  & 1.9\%  \\
16 &   405  & 1.2\%  & 1.5\%  \\
17 &   417  & 0.9\%  & 0.5\%  \\
18 &   340  & 0.7\%  & 0.6\%  \\
19 &   281  & 0.5\%  & 0.7\%  \\
20 &   218  & 0.4\%  & 0.5\%  \\
21 &   176  & 0.3\%  & 0.0\%  \\
22 &   132  & 0.2\%  & 0.8\%  \\
23 &    91  & 0.2\%  & 0.0\%  \\
24 &    72  & 0.1\%  & 0.0\%  \\
25 &    25  & 0.1\%  & 0.0\%  \\
\bottomrule
\end{tabular}
}
\end{table}

\end{document}